\documentclass{emulateapj}
\usepackage{apjfonts,mathptmx,amsmath}

\newcommand{\kB}{{k_\mathrm{B}}}
\newcommand{\mH}{{m_\mathrm{H}}}
\newcommand{\rS}{{r_\mathrm{S}}}

\newcommand{\Nth}{{N_\mathrm{th}}}
\newcommand{\Npl}{{N_\mathrm{pl}}}
\newcommand{\Nc}{{N_\mathrm{c}}}

\newcommand{\xc}{{x_\mathrm{c}}}
\newcommand{\xb}{{x_\mathrm{b}}}
\newcommand{\gammab}{{\gamma_\mathrm{b}}}
\newcommand{\gammac}{{\gamma_\mathrm{c}}}

\newcommand{\cs}{{c_\mathrm{s}}}
\newcommand{\cf}{{c_\mathrm{f}}}
\newcommand{\cA}{{c_\mathrm{A}}}

\begin{document}

\title{MHD Simulations of Accretion onto Sgr A*: \\
  Quiescent Fluctuations, Outbursts, and Quasi-Periodicity}

\author{
  Chi-kwan Chan\altaffilmark{1},
  Siming Liu\altaffilmark{2},
  Christopher L. Fryer\altaffilmark{1,3},
  Dimitrios Psaltis\altaffilmark{1,4},\\
  Feryal \"Ozel\altaffilmark{1},
  Gabriel Rockefeller\altaffilmark{1,3},
  and Fulvio Melia\altaffilmark{1,4,5}
}

\altaffiltext{1}{Department of Physics, The University of Arizona,
                 Tucson, AZ 85721}
\altaffiltext{2}{Theoretical Division, Los Alamos National Laboratory, 
                 Los Alamos, NM 87545}
\altaffiltext{3}{Computational Computer Science Division,
                 Los Alamos National Laboratory, Los Alamos, NM 87545}
\altaffiltext{4}{Steward Observatory, The University of Arizona,
                 Tucson, AZ 85721}
\altaffiltext{5}{Sir Thomas Lyle Fellow and Miegunyah Fellow}

\begin{abstract}
  High resolution observations of Sgr A* have revealed a wide variety
  of phenomena, ranging from intense rapid flares to quasi-periodic
  oscillations, making this object an ideal system to study the
  properties of low luminosity accreting black holes.  In this paper,
  we use a pseudo-spectral algorithm to construct and evolve a
  three-dimensional magnetohydrodynamic model of the accretion disk in
  Sgr A*.  Assuming a hybrid thermal-nonthermal emission scheme, we
  show that the MHD turbulence can by itself only produce factor of
  two fluctuations in luminosity.  These amplitudes in variation
  cannot explain the magnitude of flares observed in this
  system. However, we also demonstrate that density perturbations in
  the disk do produce outbursts qualitatively similar to those
  observed by XMM-{\it Newton} in X-rays and ground-based facilities
  in the near infrared.  Quasi-periodic oscillations emerge naturally
  in the simulated lightcurves.  We attribute these to
  non-axisymmetric density perturbations that emerge as the disk
  evolves back toward its quiescent state.
\end{abstract}


\section{Introduction}
\label{sec:introduction}

Accreting black holes of all masses, from stellar-mass systems in
X-ray binaries to supermassive objects in active galactic nuclei
(AGN), are highly variable, exhibiting a wide variety of outbursts
from simple flares to relativistic jets.  Periodic or quasi-periodic
signals in these outbursts \citep{vanderKlis2006}, if connected to the
orbital period at the black hole's innermost stable circular orbit
(ISCO, also known as the marginally stable orbit), provide ideal
probes of the spacetime in that region \citep[see][]{Melia2001b,
Psaltis2003}.  However, using observations of the emission from the
ISCO as probes requires understanding the dynamics of accretion near
the black hole and the origin of these oscillations.

Although these quasi-periodic oscillations (QPOs) are well-established
observationally, particularly in stellar-mass systems, their origin is
still unclear.  Quasi-periodic variability has emerged from
theoretical work only under restrictive assumptions, either
analytically in idealized disks \citep[e.g.,][and references cited
therein]{Kato2001}, or numerically in hydrodynamic
\citep[e.g.,][]{Milsom1997} and magnetohydrodynamic \citep[MHD;
e.g.,][]{Tagger2006} simulations with large-scale magnetic fields.
However, global calculations including the effects of the
magnetorotational instability (MRI) and MRI-driven turbulence have not
yet produced quasi-periodic signals \citep[e.g.,][and references cited
therein]{Hawley2001b, Armitage2003, Arras2006}.  In addition, it is
possible that the origin of these oscillations is different for
different accreting black hole systems.

At the Galactic center, the compact radio source Sagitarius A* (Sgr
A*) derives its power from accretion onto a $\sim 3.4 \times 10^6
M_\sun$ black hole \citep{Schodel2002, Ghez2004}.  Occasionally, rapid
flares are observed from this object in the X-rays
\citep{Baganoff2001} and in the near infrared \citep[NIR;
see][]{Genzel2003}.  During these flares, there is evidence for a
quasi-periodic modulation of the X-ray emission, with a period of
$\sim 17$~mins that is comparable to the orbital period at the ISCO of a
Schwarzschild black hole with the mass of Sgr A*.

The spectra and polarization of the source during the quiescent state
of Sgr A* suggest that the long-wavelength emission is due to
synchrotron radiation, whereas the more energetic X-ray emission is
due to inverse Compton scattering and thermal bremsstrahlung
\citep[see, e.g.,][]{Melia1992, Narayan1995}.  The rapid increase in
the NIR and X-ray fluxes during the flares can then be accounted for
by a transient acceleration of the energetic electrons responsible for
the emission \citep{Markoff2001, Yuan2004} or by a rapid increase of
the accretion rate.  Such an increase can be caused either by the
highly variable nature of the accretion flow or by an external surge
of matter.  In this paper, we will address the possibility that the
X-ray flares are due to a rapid increase of the accretion rate,
including situations in which this is induced by infalling clumps of
plasma.

We will use the pseudo-spectral algorithm of \citet{Chan2005,
Chan2006a, Chan2006b} to simulate the effects of the
magnetohydrodynamic turbulence on the accretion disk that surrounds
the black hole.  We will focus on the long-wavelength emission of Sgr
A* and, therefore, consider only synchrotron emission from a hybrid
plasma \citep{Ozel2000}.  For a study of spectral properties including
synchrotron emission/absorption, free-free emission/absorption, and
Compton/inverse Compton scattering, we refer to \citet{Ohsuga2005}.
We calibrate our simulations using the quiescent spectrum of Sgr A*
and aim to simulate the dynamical evolution of the system caused
either by the MHD turbulence or by an external surge of matter.  In
agreement with \citet{Goldston2005}, we find that the MRI-driven
turbulence alone cannot produce variations in the radiation flux that
are as large as the observed flares.  We also do not find any
significant quasi-periodic oscillations during the quiescent-state
flux.

To produce the observed flares, we locally perturb the density of the
disk to simulate the effects of ``clumpy material'' raining down onto
it from the large-scale flow \citep{Falcke1997, Coker1999,
Tagger2006}. Here, again, observations help to constrain the
perturbation, particularly its location in the disk.  Most of Sgr A*'s
luminosity is emitted at the mm/sub-mm spectral excess, suggesting for
our calculation that the accretion disk extends out to 5--25
Schwarzschild radii \citep{Melia2000, Melia2001a}.  Our density
perturbations are introduced locally within the disk, rather than from
a global simulation of the infall.  However, even with this
limitation, we find that we have enough freedom to construct a density
perturbation that produces flares qualitatively similar to those
observed in Sgr A*.  The simulated flares exhibit quasi-periodic
oscillations.  If such density perturbations are indeed the cause of
black-hole outbursts, this model may be used to not only study the
properties of spacetime near Sgr A*, but also the near-horizon
environment in other low-luminosity AGNs.

In \S\ref{sec:hybrid} of this paper we present our modifications to
the hybrid thermal-nonthermal emission model of \citet{Ozel2000} by
introducing a cooling break to the nonthermal component and a more
reasonable treatment of the low energy cut-off in the electron
distribution.  In \S\ref{sec:constraints}, we focus on the
observational constraints on both our emission model and the
conditions at the outer boundary of our simulated disk.  In
\S\ref{sec:setup}, we summarize the physical setup of our simulations
based on observations.  In \S\ref{sec:quiescent}, we present our
results on the quiescent state and compare the properties of the
magnetohydrodynamic turbulent plasma with those of previous numerical
studies.  We present the results of simulations involving a density
perturbation in the disk in \S\ref{sec:flares}.  We also discuss
accretion rates, the disk morphology, and lightcurves from the
quiescent and perturbed simulations in this section.  We summarize the
limitations of our simulations in \S\ref{sec:limitations}.  We
conclude with a discussion of the broader impact of these results in
\S\ref{sec:conclusions}.

\section{Emission Model}
\label{sec:hybrid}

In this section, we first describe how to incorporate special- and
general-relativistic effects for the radiative transfer in a
pseudo-Newtonian gravity.  We then propose a hybrid thermal-nonthermal
model for the electron distribution and consider its synchrotron
emissivity from a region near the ISCO.

\subsection{Radiative Transfer in Pseudo-Newtonian Gravity}

We will use cylindrical coordinates $(r,\phi,z)$ throughout this
paper.  Because the vertical size of our computational domain is small
compared to the disk radius, we will ignore gravity in the vertical
direction.  We denote the observed frequency at infinity by $\nu_0$.
The corresponding frequency measured by a stationary observer in the
local free-falling frame in a Schwarzschild spacetime, i.e., without
gravitational redshift, is given by the transformation
\begin{equation}
  \nu' = \frac{\nu_0}{\sqrt{1 - \rS/r}}\;,
\end{equation}
where $\rS = 2 GM/c^2$ is the Schwarzschild radius, and $G$, $M$, and
$c$ are the gravitational constant, mass of the central black hole,
and speed of light, respectively.  The flux density observed at Earth
is
\begin{equation}
  F_{\nu_0} = \frac{1}{D^2}\int I'_{\nu'}
  \left(1 - \frac{\rS}{r}\right)^{3/2} dA\;, \label{eq:F_nu0}
\end{equation}
where $D \approx 8.5 \mathrm{kpc}$ is the distance to the Galactic
center and $I'_{\nu'}$ is the specific intensity measured in the local
free-falling frame.  The area element, $dA$, with general-relativistic
correction, is
\begin{equation}
  dA = \frac{\cos i}{\sqrt{1 - \rS/r}}\;r\;dr\;d\phi\;,
\end{equation}
where $i$ is the inclination angle between the disk axis (i.e., the
$z$-axis in our simulations) and the line of sight.

The specific intensity is computed in a frame comoving with the
plasma.  The transformation between the comoving frame frequency and
the local free-falling frame frequency is
\begin{equation}
  \nu = \frac{1 - \beta\cos\alpha_v'}{\sqrt{1 - \beta^2}}\;\nu'\;,
\end{equation}
and the corresponding transformation for the specific intensity is
\begin{equation}
  I_\nu =
  \left(\frac{1 - \beta\cos\alpha_v'}{\sqrt{1 - \beta^2}}\right)^3
  \;I'_{\nu'}\;,
\end{equation}
where $\cos\alpha'_v$ is the cosine of the angle between the velocity
$\mathbf{v}$ and the line of sight.  If $\mathbf{\hat{v}} = \mathbf
v/|\mathbf v|$ is the directional vector of velocity and
$\mathbf{\hat{i}} = -\mathbf{\hat{r}} \sin i \sin\phi -
\mathbf{\hat{\boldsymbol{\phi}}} \sin i \cos\phi + \mathbf{\hat{z}}
\cos i$ points to the observer from the disk, then
\begin{equation}
  \cos\alpha_v' = \mathbf{\hat{v}} \cdot \mathbf{\hat{i}}\;.
\end{equation}
Note that for azimuthal velocity we use $v_\phi = \gamma\beta c$ to
take into account special-relativistic effects in pseudo-Newtonian
gravity \citep{Abramowicz1996}, so that
\begin{equation}
  \beta = \left[\left(\frac{v_\phi}{c}\right)^2 + 1\right]^{-1/2}
\end{equation}
is always less than unity.

For simplicity, we assume a time-independent transfer equation
\begin{equation}
  \frac{dI_\nu}{dl} = j_\nu - \alpha_\nu I_\nu
\end{equation}
for each snapshot of our simulations, where $l$ is the line element
along the ray (parallel to the directional vector $\mathbf{\hat i}$),
and $j_\nu$ and $\alpha_\nu$ are the emission and absorption
coefficients, respectively.  The source function of hybrid synchrotron
emission, i.e., the ratio of the emission to the absorption
coefficient, $S_\nu=j_\nu/\alpha_\nu$, is not equal to a blackbody at
the local temperature. This is especially true at wavelengths longer
than the peak of the radio/NIR spectrum of the source
\citep{Ozel2000}.  On the other hand, near the peak of the radio/NIR
spectrum, the difference between the correct source function and the
blackbody function is negligible as long as the non-thermal electrons
are a small fraction of the thermal population \citep{Ozel2000}.
Furthermore, at even shorter wavelengths, where the emission is
optically thin, the source function does not enter the calculation.
Because evaluating the hybrid source function at every wavelength for
our particular electron distribution is a time-consuming numerical
step, we will focus our attention to wavelengths comparable or shorter
than the peak of the radio/NIR spectrum and approximate the source
function with the blackbody function.  The transfer equation, written
in terms of physical depth, takes the form
\begin{equation}
  \frac{dI_\nu}{dl} = j_\nu\left(1 - \frac{I_\nu}{B_\nu}\right)\;.
  \label{eq:dI_dl}
\end{equation}
Note that this approximation has only minor effects on the
optically-thin portion of the radiation spectrum. This equation can be
integrated numerically along rays parallel to the line of sight by the
first order forward difference equation
\begin{equation}
  I_\nu^{n+1}
  = I_\nu^{n} + j_\nu\left(1 - \frac{I_\nu^{n}}{B_\nu}\right)\Delta l
\end{equation}
through the computational domain.  The superscript $n$ denotes the
steps and $\Delta l = (\Delta z/\cos i)$ is the finite difference
``line element''.  For each ray, if $I_\nu^n$ exceeds $B_\nu$ before
integrating through the whole domain, we simply stop and set $I_\nu$
equal to the local blackbody value.

\subsection{Electron Distribution}

\begin{figure}
  \includegraphics[scale=0.75,trim=18 9 0 18]{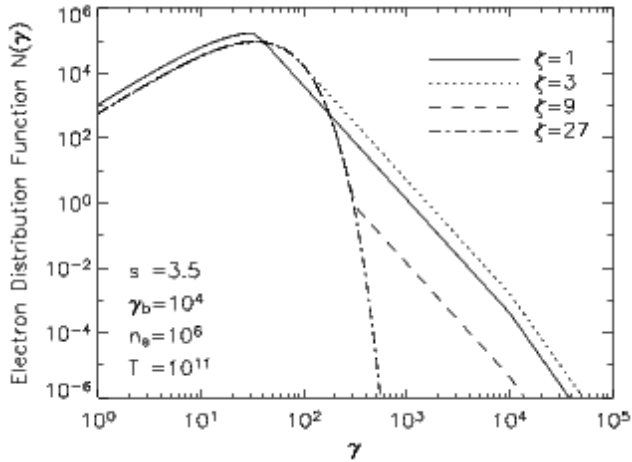}
  \caption{The electron distribution function $N(\gamma)$ for
    different choices of $\zeta$.  The other model parameters are
    fixed as shown in the legend. At the cooling break $\gammab$, the
    power-law index increases by 1.}
  \label{fig:N_gamma}
  \vspace{6pt}
\end{figure}

\citet{Ozel2000} proposed a hybrid thermal-nonthermal model for
synchrotron radiation in low-luminosity AGN.  When applied to Sgr A*,
the hybrid model shows better agreement with observations compared to
a pure thermal model.  However, recent polarization measurements of
Sgr A* show that the rotation measure in the emission region is small,
which suggests that the low-energy component of electrons is
relativistic and likely thermal \citep{Marrone2006}.  We modify the
hybrid model so that the nonthermal distribution contributes only at
high energies.  That is, for some critical value $\gammac$, we assume
\begin{equation}
  N(\gamma) = \left\{\begin{array}{lll} \Nth(\gamma) & , & \gamma <
    \gammac\;, \\ \Npl(\gamma) & , & \gamma \ge \gammac\;,
  \end{array}\right.
\end{equation}
where $\gamma$ is proportional to the (ultra-relativistic) electron
energy.  When most of the electrons are thermalized, our distribution
function is a close approximation to the hybrid model of
\citet{Ozel2000}.  Otherwise, our modified hybrid model keeps the
low-energy electrons in a thermal distribution.

In order to obtain analytical expressions for the normalization, we
take the domain of $\gamma$ to be all positive real numbers, and use
the ultra-relativistic Maxwell-Boltzmann distribution
\begin{equation}
  \Nth(\gamma) = \Nc \frac{\gamma^2}{\gammac^2}
  \exp\left(\frac{\gammac - \gamma}{\theta_e}\right),
\end{equation}
where $\Nc$ is the normalization and $\theta_e = \kB T_e / m_e c^2$ is
the dimensionless electron temperature.  We use $\kB$ to denote the
Boltzmann constant, $T_e$ to denote the electron temperature, which is
assumed to be equal to the plasma temperature $T$, and $m_e$ to denote
the electron mass.  Provided $\theta_e \gg 1$, the ultra-relativistic
approximation will not introduce an appreciable error.  Note that the
maximum of the thermal distribution is located at $\gamma =
2\theta_e$.  It is convenient to introduce a parameter
\begin{equation}
  \zeta = \gammac/2\theta_e \ge 1\;,
\end{equation}
and specify it in our model.

We also introduce an additional parameter for the cooling break.  The
broken power law is given by
\begin{equation}
  \Npl(\gamma) = \Nc \left\{\begin{array}{lll}
    \ \ \ \ \gammac^s/\gamma^s     & , & \gamma <   \gammab\;, \\
    \gammab \gammac^s/\gamma^{s+1} & , & \gamma \ge \gammab\;.
  \end{array}\right.
\end{equation}
The power-law index $s$ describes the spectrum of injected electrons
and is believed to be greater than unity.  Because the synchrotron
cooling timescale is proportional to $\gamma^2$, the electrons in the
high-energy tail cool more rapidly.  The parameter $\gammab$ therefore
controls the location of this cooling break.

The symbol $\Nc$ in the above equations denotes the normalization of
the distribution at $\gammac$.  Integrating the distribution, it is
easy to show that
\begin{eqnarray}
  \Nc & = & \frac{\rho}{\mH} \left\{
    \frac{2\theta_e^3}{\gammac^2}
    \left[\exp\left(\frac{\gammac}{\theta_e}\right)-1\right] - 
    \frac{2\theta_e^2}{\gammac} - \theta_e +
  \right. \nonumber\\
  & & \ \ \ \ \ \ \ \ \ \left.
    \frac{1}{s-1}\left[\gammac -
    \gammab\left(\frac{\gammac}{\gammab}\right)^s\right] +
    \frac{1}{s}\gammab\left(\frac{\gammac}{\gammab}\right)^s
  \right\}^{-1},
\end{eqnarray}
where $\rho$ is the mass density and $\mH$ is the mass of hydrogen.

Figure~\ref{fig:N_gamma} shows the electron distribution function
$N(\gamma)$ for different choices of $\zeta$ with fixed density $n_e =
\rho/m_{\rm H}$ and temperature.  The function approaches a pure
thermal distribution as $\zeta \rightarrow \infty$.  As we decrease
$\zeta$, more and more electrons are placed in the nonthermal tail.
Note that the spectral index of the power-law component increases by
one at the location of the cooling break $\gammab$.  This modified
hybrid thermal-nonthermal model will be used to fit the spectrum of
Sgr A* up to the X-ray band in \S\ref{sec:constraints}.

\subsection{Synchrotron Radiation}

\begin{figure*}
  \includegraphics[scale=0.75,trim=9 9 0 27]{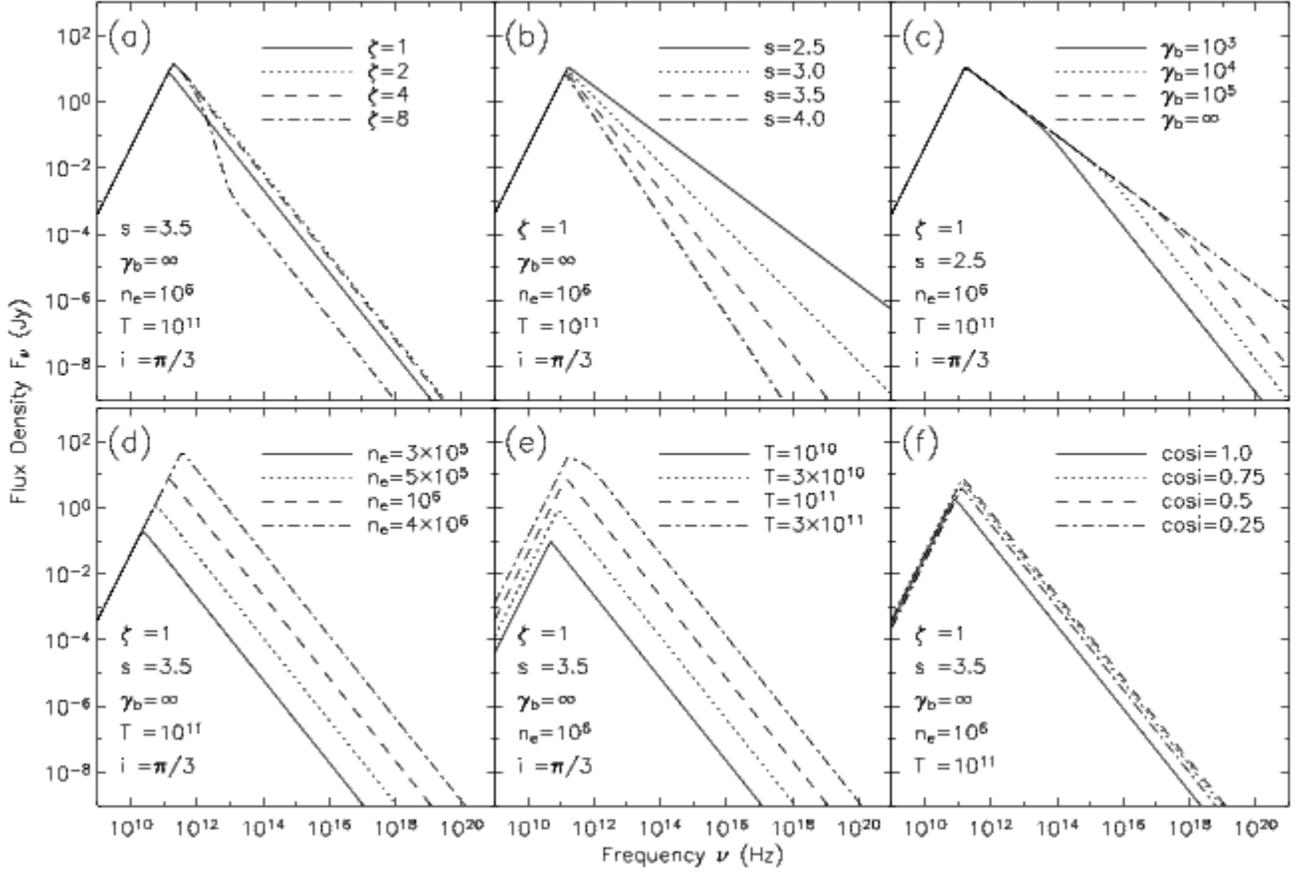}
  \caption{Dependence on the model parameters of the flux density
    $F_\nu$ observed at Earth. The flux densities are computed from
    our emission model assuming a simplified disk.  The top panels
    [from (a) to (c)] present the dependence on different parameters
    in the emission model, while the bottom panels present the
    dependence on the parameters of the disk itself.  The units for
    $T$ are $\mathrm{K}$; the units for $n_e$ are $\mathrm{cm}^{-3}$.
    The non-monotonic dependence of the spectrum on $\cos i$ is caused
    by the combined effects of Doppler boosting, projection, and the
    pitch-angle dependence of the synchrotron radiation.}
  \label{fig:comp}
  \vspace{6pt}
\end{figure*}

The emission coefficient in equation~(\ref{eq:dI_dl}) is given by
\begin{eqnarray}
  j_\nu & = & \frac{\sqrt{3}e^2}{2c}\nu_B \sin\alpha_B
  \int_0^\infty N(\gamma) F(x) d\gamma\;, \label{eq:j_nu_general}
\end{eqnarray}
where $\nu_B = e B/2\pi m_e c$ is the non-relativistic cyclotron
frequency, and $\alpha_B$ is the angle between the magnetic field and
the line of sight in the comoving frame.  Here we use $e$ to denote
the electron charge and $B$ to denote the magnitude of the magnetic
field.  The function $F(x)$ is defined by
\begin{equation}
  F(x) = x \int_x^\infty K_{5/3}(z) dz
\end{equation}
\citep{Pacholczyk1970}, where $x$ is related to $\gamma$ according to
$x = x_0/\gamma^2$, with
\begin{equation}
  x_0 =  \frac{2\nu}{3\nu_B\sin\alpha_B}\;.
\end{equation}
We assume that the sub-grid fluctuation of the magnetic field is
small.  Writing $\mathbf{\hat{B}} = \mathbf{B}/|\mathbf{B}|$, we
simply use
\begin{equation}
  \cos\alpha_B' = \mathbf{\hat{B}}\cdot\mathbf{\hat{i}}
\end{equation}
to obtain the angle between the magnetic field and the line of sight.
The comoving angle cosine is then given by the transformation
\begin{equation}
  \cos\alpha_B = \frac{\cos\alpha_B' - \beta}{1 - \beta\cos\alpha_B'}\;.
\end{equation}

Using $x_M = x_0/\theta_e^2$, the emission coefficient can be written
as
\begin{eqnarray}
  j_\nu & = & \frac{e^2 \nu}{\sqrt{3}c} \Nc \left[
    \frac{e^{2\zeta}}{4\zeta^2 \theta_e}I_{2\zeta}(x_M) + 
    \frac{\gammac^s}{2} J_{s,\gammab}(x_0,\gammac)
  \right]\;. \label{eq:j_nu}
\end{eqnarray}
The partial thermal synchrotron function, $I_{2\zeta}(x_M)$, is given
by the integral
\begin{equation}
  I_{2\zeta}(x_M) = \frac{1}{x_M}
    \int_0^{2\zeta} z^2 e^{-z} F\left(\frac{x_M}{z^2}\right) dz\;,
\end{equation}
where $z = \gamma/\theta_e$.  It does not have any known analytical
form.  We approximate it by a piecewise power law, i.e., for each
fixed $\zeta$, we pre-compute the function numerically as a lookup
table and carry out linear interpolations on a log-log scale.
Similarly, the function $J_{s,\gammab}(x_0,\gammac)$ is given by
\begin{eqnarray}
  J_{s,\gammab}(x_0,\gammac) & = &
  \frac{1}{x_0^{(s+1)/2}} \int_\xb^\xc x^{(s-3)/2} F(x) dx + \nonumber\\
  & & \frac{\gammab}{x_0^{(s+2)/2}} \int_0^\xb x^{(s-2)/2} F(x) dx\;,
  \label{eq:J}
\end{eqnarray}
where $\xc = x_0/\gammac^2$ and $\xb = x_0/\gammab^2$.  The integrals
are related to a class of hypergeometric functions.  However, because
there is no convergent algorithm to compute this specific class of
hypergeometric functions, we will simply perform the integral
numerically, for each $s$ and $\gammab$.  We use a piecewise power law
to approximate the first integral as a function of $\xc$.

Figure~\ref{fig:comp} demonstrates how the spectrum depends on the
different parameters. To highlight the characteristics of the emission
model, we assume for illustrative purpose that the disk is uniform and
rotates with a constant (dimensionless) azimuthal velocity $\beta =
2/3$. We also assume an azimuthal magnetic field, whose energy density
is $10\%$ of the internal energy of the gas, and use a disk volume of
$4\pi \times 10^{38} \mathrm{cm}^3$, which is of the same order as our
computational domain (see \S\ref{sec:setup}).  In the first row of
panels, we vary different parameters of the electron distribution.  It
is interesting to note that in panel (a), the nonthermal tail depends
only weakly on $\zeta$ when $\zeta \lesssim 4$.  Panel (b) shows how
the spectrum depends on the power-law index $s$.  Indeed, from
equations~(\ref{eq:j_nu}) and (\ref{eq:J}), it can be seen immediately
that the nonthermal tail has a broken power-law spectrum with indexes
$(s-1)/2$ and $s/2$. Panel (c) shows the effect of changing $\gammab$.

The second row shows the dependence of the spectrum on $n_e$, the
temperature $T$ (note that we assume that the electron temperature is
equal to the ion temperature), and the inclination angle of the disk
$i$.  As the density is increased, the disk becomes optically thin at
higher frequencies [see panel (d)].  The optically-thick portion of
the spectrum is blackbody-like and is proportional to the projected
source size and the temperature.  Increasing the temperature will
increase the blackbody flux linearly.  The emission efficiency in the
optically thin portion is proportional to the square of the
temperature, which explains the change in shape of the spectrum at $T
= 3 \times 10^{11}\mathrm{K}$ [see panel (e)].  Panel (f) shows the
dependence on the inclination angle $i$ of the disk.  On one hand, a
larger value of $i$ results in a smaller projected source size and a
small angle between the magnetic field and the line of sight at the
blue-shifted side of the disk, implying less efficient synchrotron
radiation.  On the other hand, a larger value of $i$ enhances the
effect of Doppler boosting.  We found that the highest peak frequency
is reached when $\cos i= 1/2$, which, obviously, depends on the
azimuthal velocity we adopt.

\section{Observational Constraints}
\label{sec:constraints}

\begin{figure}
  \includegraphics[scale=0.75,trim=9 9 0 18]{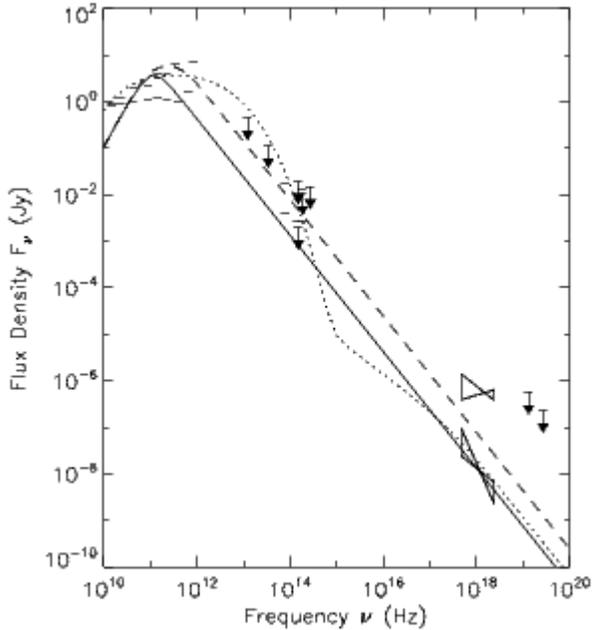}
  \caption{Model fit to the observed broadband spectrum of Sgr A*.
    The horizontal bars show only the extrema obtained from different
    observations, while the arrows at NIR and $\gamma$-ray wavelengths
    give upper bounds on the flux densities in the quiescent state.
    We use $T = 10^{11} \mathrm{K}$ and $n_e = 5 \times 10^6
    \mathrm{cm}^3$ at the outer boundary of the disk.  The solid line
    is the spectrum computed from our emission model with $\zeta = 1$,
    $s=3.5$, $\gammab = \infty$, and $\cos i = 1/2$ from a snapshot at
    the $t = 13$~hr of the quiescent simulation. The dashed line
    corresponds to the perturbed simulation.  To demonstrate that this
    is not a unique fit, we compute the quiescent spectrum with a
    temperature that is 10 times higher throughout the disk and
    $\zeta=10$, $s=2.5$ and $\gammab=10^5$ (the other parameters are
    the same as the solid line); the result is shown by the dotted
    line.}
  \label{fig:spect}
  \vspace{6pt}
\end{figure}

\begin{figure*}
  \includegraphics[scale=0.75,trim=18 9 9 9]{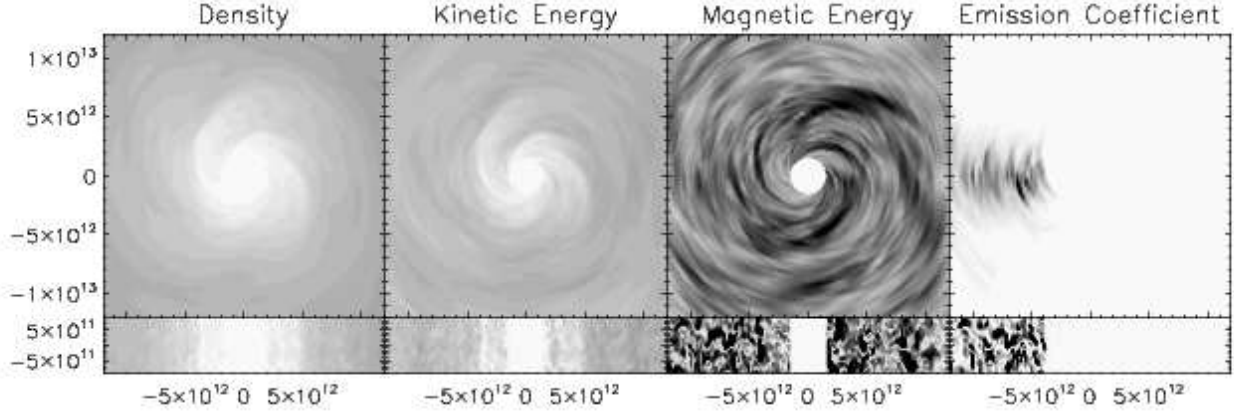}
  \caption{Gray-scale plots of different quantities in the unperturbed
    simulation at late time ($t=13$~hr).  From left to right the
    panels show the density $\rho$, kinetic energy density $\rho
    v^2/2$, magnetic energy density $B^2/8\pi$, and the emissivity in
    the infrared ($\nu_0=1.8\times10^{14}\mathrm{Hz}$) band with
    relativistic corrections ($\cos i=1/2$ in this case, see text for
    details). The polar plots in the top row are vertically-averaged
    results, while the rectangular plots along the bottom show the
    variation with height in the disk. The asymmetry in the effective
    emission coefficient comes from the Doppler effect.}
  \label{fig:cont_unp}
  \vspace{6pt}
\end{figure*}

Figure~\ref{fig:spect} shows the observed flux of Sgr A* at different
frequencies \citep[see][for the instruments and references]{Liu2004}.
The horizontal bars show the maximum and minimum fluxes observed so
far at the corresponding frequencies.  The arrows at NIR and
$\gamma$-ray wavelengths give the upper bounds for the quiescent state
emission.  Fitting these data, one obtains a (non-unique) set of
parameters for the emission model and physical conditions of the
accretion disk.

The mm/submm spectrum of Sgr A* has been modeled in different ways by
various groups.  \citet{Narayan1995} used thermal emission from an
advection-dominated accretion flow.  \citet{Mahadevan1998} and
\citet{Ozel2000} incorporated the effects of a hybrid electron
distribution in the same model.  \citet{Melia2001c} considered thermal
synchrotron emission from a compact accretion disk, whereas
\citet{Markoff2001} used the emission of a thermal jet.  For a
magnetic field energy density below equipartition with the energy
density of the plasma as suggested by MHD simulations
\citep{Hawley2000}, these studies generally point to an emission
region of a few Schwarzschild radii with a plasma temperature of $\sim
10^{11}\mathrm{K}$ and a density of $\sim 10^6 \mathrm{cm}^{-3}$.

Given the low luminosity of Sgr A*, the accretion flow is radiatively
inefficient \citep{Melia1992, Melia1994, Narayan1995, Yuan2003,
Yuan2004}.  The temperature of the plasma therefore should be close to
its virial value.  As mentioned above, we assume for simplicity that
the electron and proton temperatures are equal.  A realistic treatment
of the thermal coupling between the protons and the electrons is not
feasible in our time-dependent simulations.  For a given outer
boundary radius of the accretion disk, the scale height of the disk
can be estimated from the virial temperature, and we have only the gas
density left as a free parameter.

We take the outer boundary of the accretion disk to be $21.5 \rS$ in
our simulations.  We can therefore use the broadband spectrum of Sgr
A* to constrain the density at the outer boundary, which determines
the density normalization of the whole disk, the inclination angle of
the disk, and other parameters describing the nonthermal population of
the electrons (specifically, $\zeta$, $s$, and $\gammab$).

The solid line in Figure~\ref{fig:spect} is an emission spectrum
computed from a late time snapshot (at $t=13$~hr) of our
quiescent-state simulation (see \S\ \ref{sec:quiescent} for details)
with $T=10^{11} \mathrm{K}$ and $n_e = 10^6 \mathrm{cm}^{-3}$ at the
outer boundary.  The disk inclination angle is $\cos i = 1/2$, while
the emission model parameters are $\zeta = 1$, $s=3.5$, and $\gammab =
\infty$.  We note that the spectrum peaks at $\sim 1.2 \times
10^{11}$~Hz with a peak flux density of $\sim 3$~Jy.  Observations
suggest that Sgr A*'s spectrum peaks at $220$~GHz, slightly above the
model predicted value, with a flux density of $\sim 3$~Jy (Marrone et
al. 2006).  In order to better understand this difference, we sampled
the parameter space to improve the quality of the fitting.  For the
chosen snapshot of our quiescent-state simulation, we either
over-predict the peak flux density or have a lower peak frequency as
shown in the figure.  These deviations suggest that the real electron
temperature might be lower than the plasma temperature, as also
expected on theoretical grounds, while the electron density might be
higher than in our calculations; this would shift the simulated
emission peak frequency and flux density closer to the observed
values.  However, quantitative fitting to the spectrum is beyond the
scope of this paper, which focuses on the variability of the NIR and
X-ray emission.  We will adopt the above model parameters in the
following sections.

Note that, with $\gammab=\infty$, we only need two parameters to
describe the spectrum of nonthermal electrons.  Here we neglect other
radiation processes that may dominate in the X-rays (i.e.,
bremsstrahlung and comptonization).  Moreover, at least part of the
quiescent-state X-ray emission from Sgr A* is produced near the
capture radius of the accretion flow by a thermal plasma
\citep{Baganoff2001, Baganoff2003}.  As a result, the observed X-ray
flux needs to be treated as an upper limit for the X-ray emission from
our model of the small accretion disk.  The above model fitting
therefore suggests that for $\zeta = 1$, the power-law index can not
be less than 3.5.  For large values of $\zeta$, one may adopt a small
$s$ without violating the X-ray upper limit.  These models, however,
will predict very weak NIR emission.  The parameter $\gammab$ needs to
be included to bring the NIR flux close to the observed values during
flares.

The dashed line in Figure~\ref{fig:spect} corresponds to the perturbed
simulation discussed in \S\ref{sec:flares}, which predicts an X-ray
flare with a soft spectrum and a flux density 8 times higher than the
quiescent level.  Further analysis of small flares observed by the
\emph{Chandra} and \emph{XMM}-Newton telescopes are needed to test
this prediction.  As we pointed out above, the values of model
parameters presented earlier do not constitute a unique fit to the
spectrum.  The dotted line in the figure is another possible fit to
the spectrum\footnote{The spectrum is computed by increasing the
temperature by a factor of ten in our simulation.  If we had used
Newtonian gravity in our calculation, we could simply rescale the unit
of velocity to obtain the answer for a different temperature.
However, pseudo-Newtonian gravity (or full general relativity) sets
the characteristic velocity to the speed of light.  We cannot rescale
the temperature without rerunning the simulation.  The fit shown here,
therefore, is only a quantitative demonstration of the uncertainty in
our emission model.}.  The corresponding model parameters are $\zeta =
10$, $s=2.5$, $\gammab = 10^5$, and $T = 10^{12}\mathrm{K}$.  The
values of the other parameters are the same as those used to produce
the solid line.

\section{Physical Setup}
\label{sec:setup}

We take the mass of Sgr A* to be $M = 3.4\times10^6 M_\sun$
\citep{Schodel2003, Ghez2003}.  From the physical conditions in the
disk estimated in the previous section, the density and temperature at
the outer boundary ($r=21.5\ r_S$) are fixed at $\rho_0 \approx 1.7
\times 10^{-18} \mathrm{g\;cm}^{-3}$ and $T_0 = 10^{11} \mathrm{K}$.
At this temperature, the plasma is fully ionized, so we take the mean
molecular weight to be $\mu = 0.5$.  We also assume the MHD equations
are valid although the electron distribution has a nonthermal
component.  An absorbing layer is placed between $20 \rS < r < 21.5
\rS$ to attenuate waves reflecting back into the domain.

In order to focus on the effects of magnetohydrodynamic turbulence
produced by shearing, we assume a slab geometry and neglect the
vertical component of gravity.  Because the disk's mass is negligible
compared to that of the black hole, we also neglect self-gravity.
Inflowing (to the black hole) conditions are imposed at the inner
boundary at $r=1.5\,\rS$, which lies well below the ISCO ($\approx
3\rS$) in a Schwarzschild geometry.

At such high temperatures, the electrical conductivity within the
plasma is high enough that we can neglect any resistive deviations
from ideal MHD \citep[as was done, for example, in the global
simulations of][following the suggestion by Shvartsman
1971]{Igumenshchev2002}.  We also neglect molecular viscosity, which
is expected to be insignificant compared to the turbulent viscosity
arising from Maxwell (and Reynolds) stresses in the magnetized plasma.
We take the initial velocity to be Keplerian in a pseudo-Newtonian
gravitational potential, $v_{\phi0} = \sqrt{G M r}/(r - \rS)$.
Shearing leads to the development of the MRI \citep{Balbus1991}.

We start with a uniform density and temperature and an initial
magnetic field $\mathbf B = B_0\mathbf{\hat z}$ with $B_0 = 0.3
\mathrm{G}$.  A random perturbation at the $1\%$ level is added to the
temperature, and hence pressure, in order to initiate the MRI.  The
disk is allowed to evolve towards a quasi-steady state.  Note that a
linear mode analysis shows that the modes become stable when $k_z^2 >
3 (\Omega/v_{A,z})^2$, so it is important that the computational
domain is large enough to enclose unstable modes for which $\lambda_z
> B_z (r-\rS)\sqrt{\pi r/3GM\rho}$.  For our model, the critical value
is roughly $0.01 r_S$ at the inner boundary, and $1 r_S$ at the outer
boundary.  Therefore, we choose the vertical domain to be $-r_S \le z
< r_S$ to ensure that the MRI can develop over the whole disk.  We use
$257\times64\times32$ grid points in our simulations.  The vertical
resolution can resolve the most unstable wavelength outside the ISCO.
The radial and azimuthal grid points are chosen so that these two
directions are resolved as accurately as the $z$ direction.

\section{Quiescent State}
\label{sec:quiescent}

We carry out the calculation described in \S\ref{sec:setup} for 24
simulated hours, which corresponds to about 80 orbits at the ISCO.
The material inside the ISCO is accreted very quickly, in under an
hour.  Magnetohydrodynamic turbulence then kicks in because of the
MRI.  The inner disk rotates much more rapidly so turbulence is fully
developed first in the inner region; the MRI in the outer region grows
more slowly.  Turbulence is developed through the whole disk by
$t=12$~hr (note that the period at the outer boundary is around 6
hours), and the disk reaches a quasi-steady state thereafter.

\subsection{Quasi-Steady State Solution}

To illustrate the turbulent flow in the disk, we show in
Figure~\ref{fig:cont_unp} gray-scale plots of the density $\rho$,
kinetic energy density $\rho v^2/2$, magnetic energy density
$B^2/8\pi$, and the NIR (i.e., $\nu_0=1.8\times10^{14} \mathrm{Hz}$)
emissivity.  The plots show the disk profile at $t=13$~hr of the
simulation, after steady state has been reached.  The NIR emissivity
is computed with relativistic corrections, using the equation
\begin{equation}
  j_{\nu0} = \left(1-\frac{\rS}{r}\right)^2
  \left(\frac{\sqrt{1-\beta^2}}{1-\beta\cos\alpha'_v}\right)^3 j_\nu\;,
\end{equation}
where we have assumed a $60^\circ$ inclination angle for the disk.
The polar plots in the top panels are vertical-averaged quantities.
The rectangular plots along the bottom show the vertical structure at
the angles $\phi = 0$ and $\pi$.  Note that the $z$-axis is rescaled
to show the vertical structure; the disk itself is much thinner than
it appears in these plots.

Because of the high temperature, the density (and the thermal energy,
which is not shown in the plots) is rather smooth.  The region within
$3\rS$ is almost empty. Although both the density and kinetic energy
decrease inside the ISCO, there is no significant drop in the magnetic
field or the temperature there.  A weak two-armed spiral pattern, or
$m=2$ mode, appears in the density (and thermal energy).  This pattern
also shows up in the kinetic energy density plot, which is correlated
to that of the magnetic field.  The turbulent magnetic field shows
smaller structure close to the inner disk and larger eddies in the
outer region.  This agrees with the standard stability criterion of
\citet{Balbus1991}.

The strong asymmetry of the emission coefficient comes from the
Doppler shift.  Because the plasma moves in a counter-clockwise
direction (in the polar plots), the emission from the left side is
blue-shifted.  The hybrid emission spectrum decreases for frequencies
$\nu > 10^{11} \mathrm{Hz}$.  Hence, blue-shifting raises the emission
spectrum and makes the left side of the disk much brighter.  In the
rectangular plots, the patchy structure of the emission coefficient
for $r > 4 \rS$ is well correlated with the magnetic field because
$j_\nu \propto B^2$.  For the region inside $4 \rS$, the low density
and low thermal energy, together with the gravitational redshift,
reduce the emission coefficient significantly.

\subsection{Saturated Turbulent Flow}

\begin{figure}
  \includegraphics[scale=0.75,trim=18 9 0 18]{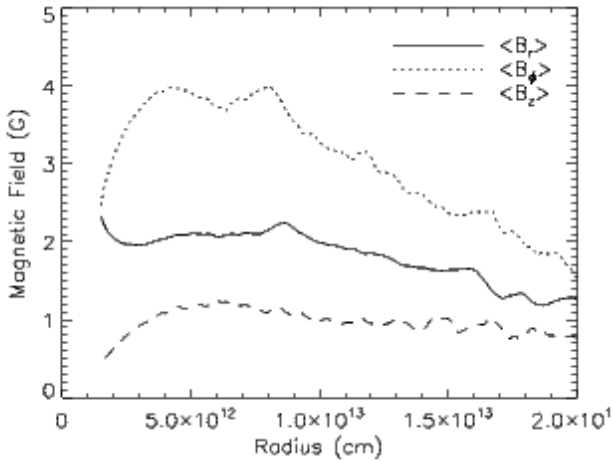}
  \caption{The root-mean-square of the different components of the
    magnetic field.  Note that, although we start with a seed magnetic
    field of 0.3~G in the $z$-direction, the magnetic field saturates
    at a value $\sim10$ times higher in the (dominant) $B_\phi$
    component.}
  \label{fig:magnetic_field}
  \vspace{6pt}
\end{figure}

\begin{figure}
  \includegraphics[scale=0.75,trim=18 9 0 18]{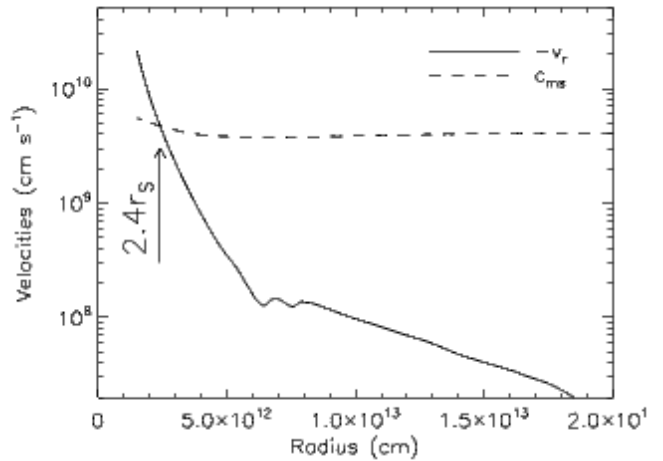}
  \caption{Average radial velocity (solid) and magnetosonic speed
    (dashed) profiles through the disk.  The intersection of the two
    lines defines the magnetosonic point $r_\mathrm{ms} \simeq 2.4
    \rS$.  The fluid inside this point loses causal contact with the
    outer part of the disk.}
  \label{fig:magnetosonic}
  \vspace{6pt}
\end{figure}

In Figure~\ref{fig:magnetic_field}, we compute the root-mean-square of
different components of the magnetic field
\begin{equation}
  \langle B_i\rangle_\mathrm{rms} =
  \left(\frac{\int B_i^2 r\;d\phi\;dz}{\int r\;d\phi\;dz}\right)^{1/2}.
\end{equation}
Although our initial magnetic field has only a vertical component with
a seed value of 0.3~G, the dominant component in the quasi-steady
state is $B_\phi$. This result is in agreement with earlier global
simulations \citep[see][]{Hawley2001a}.  The magnitude of the magnetic
field (which is dominated by the $\phi$-component) ranges from 2~G
near the outer boundary, to the peak value of 4~G near $6\rS$, and
decreases to about 2.5~G within the ISCO. The magnetic field strength
is around 10\% of the equipartition value, consistent with the results
of \citet{Hawley2001b, Hawley2002b}.

Defining an average along the $\phi$- and $z$-directions as
\begin{equation}
  \langle f \rangle_{\phi z} = \frac{\int f r\;d\phi\;dz}{\int r\;d\phi\;dz}\;,
\end{equation}
the magnetic flux along the radial direction is then given by
\begin{equation}
  \Phi_{B_r}(r) = \langle B_r \rangle_{\phi z} \int r d\phi dz\;.
\end{equation}
This should vanish for all radii because of the divergenceless
condition of the magnetic field $\nabla\cdot\mathbf B = 0$ and our
periodic $z$-axis boundary condition.  We compute $\langle B_r
\rangle_{\phi z}$ from our simulation and obtain $\langle B_r
\rangle_{\phi z}\lesssim 10^{-15}\;\mathrm{G}$ at all times.  Compared
to the root-mean-square values of the different components of the
magnetic field, this shows that our algorithm is able to preserve
$\nabla\cdot\mathbf B = 0$ down to machine accuracy.  This property is
a useful test of our simulation.

The inner edge of the accretion disk should therefore be defined as
the location where the infalling material loses causal contact with
the rest of the disk.  Hence, it may be functionally defined to be the
magnetosonic point, at which the radial velocity matches the (fast)
magnetosonic speed $\cf = \sqrt{\cs^2 + \cA^2}$, where $\cs$ is the
sound speed and $\cA = B/\sqrt{4\pi\rho}$ is the Alfv\'en speed.
Below this radius, the material loses causal contact with the rest of
the disk at larger radii.  We plot in Figure~\ref{fig:magnetosonic}
the root-mean-square of the radial velocity and magnetosonic speed as
functions of radius.  The intersection indicates the location of the
magnetosonic point, which is around $2.4\rS$.  The Keplerian period at
this radius is 10.76~min, in contrast to the 17.19~min period at the
ISCO.

\section{Flares}
\label{sec:flares}

\begin{figure*}
  \includegraphics[scale=0.75,trim=18 9 9 9]{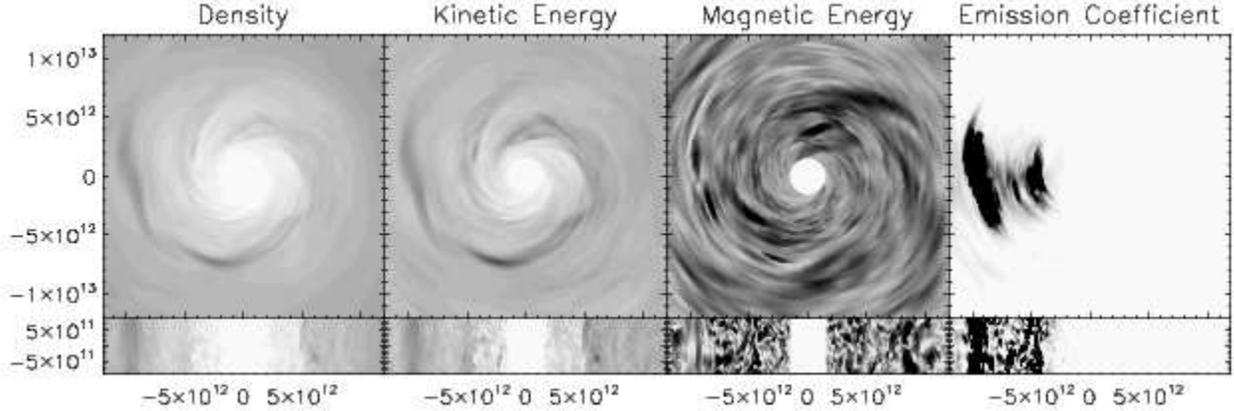}
  \caption{Same as Figure \ref{fig:cont_unp} but for the
    perturbed simulation.}
  \label{fig:cont_per}
  \vspace{6pt}
\end{figure*}

Earlier hydrodynamic and MHD simulations of the large-scale ($r \gg 20
\rS$) accretion flow \citep[e.g.][]{Ruffert1994, Coker1997,
Igumenshchev2002, Igumenshchev2003, Cuadra2005} have indicated that
the inflow onto Sgr A* is not smooth.  The time scales of these
simulations range from 1 to $10^4$ years.  Given the low-density
conditions in the surrounding medium, only parcels of plasma with
relatively small specific angular momentum find their way to small
radii \citep{Coker1997}.  The lack of strong accretion apparently
inhibits the formation of a large continuous disk extending out to the
Bondi-Hoyle capture radius (around $10^5\rS$).  Instead, the compact
disk in Sgr A* appears to be accreting clumps of plasma that ``rain''
inwards from all directions \citep{Falcke1997, Coker1999, Tagger2006}.

Each clump circularizes at a radius corresponding to its specific
angular momentum.  Some clumps presumably reach as far in as the ISCO;
most of the others probably merge with the disk at larger radii.  To
model the impact of such a clump falling onto our quasi-steady inner
disk, we simulate its effect on the luminosity of the disk by
introducing a density perturbation in the saturated disk.

\subsection{Density Perturbation}

The perturbation is introduced into the quiescent simulation after
quasi-steady state is reached, at $t = 12$~hr.  The perturbation is
Gaussian in density
\begin{eqnarray}
  \delta\rho = 2 \rho_0
  \exp\left(-\frac{r^2 + r_0^2 + 2r r_0\cos\phi}{2\sigma}\right)\;,
\end{eqnarray}
where $\rho_0 = 10^6 \mH\;\mathrm{cm}^{-3}$, $r_0 = 6 \rS$ and $\sigma
= 2\rS$.  We assume that the perturbation has a temperature of only
$10^{10}\mathrm{K}$, much cooler than the disk, to avoid a strong
increase in pressure.  This allows the clump to move with the plasma
in the disk instead of propagating out as sound waves.  The internal
energy, therefore, is raised by
\begin{equation}
  \delta E = \delta\rho \frac{3\kB}{2\mu\mH} \times 10^{10}\mathrm{K}\;.
\end{equation}
For simplicity, we also assume the clump has zero momentum.  Hence the
velocity is slowed down by
\begin{equation}
  \delta\mathbf{v} = \frac{-\delta\rho}{\rho + \delta\rho}\;\mathbf{v}\;.
\end{equation}

The simulation is carried out for 10 hours, until the accretion rate
becomes comparable to that of the quiescent state again. In
Figure~\ref{fig:cont_per}, we plot the density, kinetic energy
density, magnetic energy density, and the effective NIR emission
coefficient at $t=13$~hr.  The various physical quantities are shown
at the same time and with the same gray-scale as in
Figure~\ref{fig:cont_unp} for comparison.  The perturbation is sheared
out and forms a one-armed spiral pattern in the disk, which also
raises the kinetic energy.  The correlation between the magnetic
energy density and the gas density is less clear in the plot.  The
perturbation seems to produce a few strong magnetic spots, but it also
weakens the $m = 2$ mode.  There is a prominent feature in the
effective emission coefficient; this ``hot spot'' arises due to both
the density and the temperature enhancements.

\subsection{Accretion Rate}

\begin{figure*}
  \includegraphics[scale=0.75,trim=0 9 0 9]{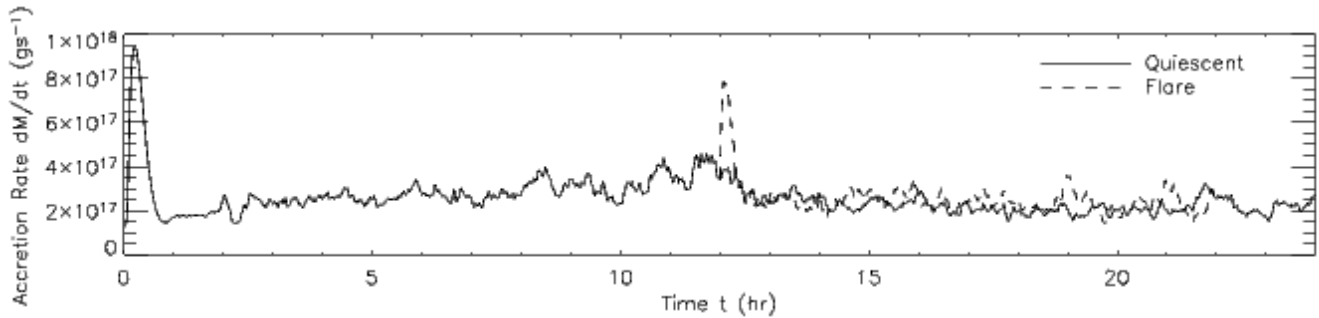}
  \caption{The variation of the accretion rate $\dot{M}$ at $r=3\rS$
    with time.  The solid line corresponds to the unperturbed
    simulation.  The peak that appears in the first hour is due to our
    initial condition of a uniform disk.  Matter below the ISCO falls
    into the black hole very quickly after the simulation begins.  The
    dashed line after $t = 12\mathrm{hr}$ is the result from the
    simulation with a density perturbation (see \S\ref{sec:flares}).
    The accretion rate rises for about half an hour, then falls back
    to the quasi-steady level.}
  \label{fig:mdot}
  \vspace{6pt}
\end{figure*}

Before looking for quasi-periodic signals in the lightcurve, we first
study variations in the accretion rate.  In Figure~\ref{fig:mdot} we
plot the accretion rate of our simulation at $r = 3\rS$ as a function
of time.  The solid line corresponds to the unperturbed simulation,
and the dashed line corresponds to the accretion rate after we
introduce the density perturbation.  The strong peak in the first hour
is due to our initial condition of a uniform disk.  The material
inside the ISCO falls into the black hole immediately after the
simulation begins.  Although the accretion rate never settles to a
constant value, it converges to a value around $2\times10^{17}
\mathrm{g\;s}^{-1}$, consistent with the value inferred from earlier
semi-analytic treatments.  The weaker peak appearing right after
$t=12$~hr is caused by the density perturbation.

Although we do not show it in the plot, the accretion rate $\dot{M}$
at radii $r > 6\rS$ oscillates between positive and negative values
(it actually ranges from $-6 \times 10^{17}$ to $1 \times 10^{18}
\mathrm{g\;s}^{-1}$).  These oscillations in $\dot M$ have
well-defined periods of about 1 hour, though they are not related to
the orbital period. They are simply pressure-driven.

\subsection{Disk Images}

Assuming an inclination angle of $60^\circ$ and using the parameters
obtained in \S\ref{sec:constraints}, we integrate the transfer
equation along parallel rays and obtain ray-traced images of the
accretion disk around Sgr A*. Compared to the emissivity plots in
Figures~\ref{fig:cont_unp} and \ref{fig:cont_per}, these images
represent the observation more directly because they are numerical
solutions of the radiative transfer equation.  They also take into
account the projection effects of the disk.

\begin{figure*}
  \includegraphics[scale=0.75]{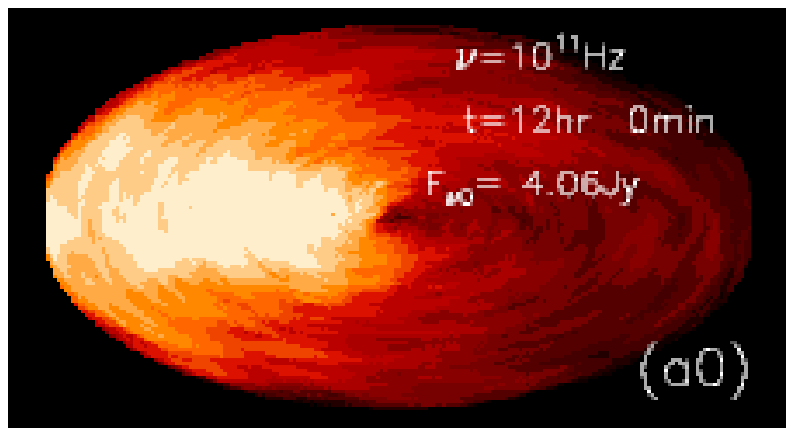}
  \includegraphics[scale=0.75]{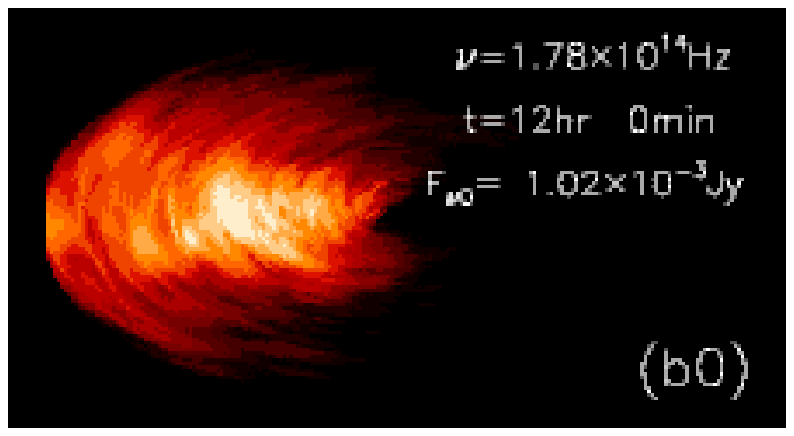}
  \includegraphics[scale=0.75]{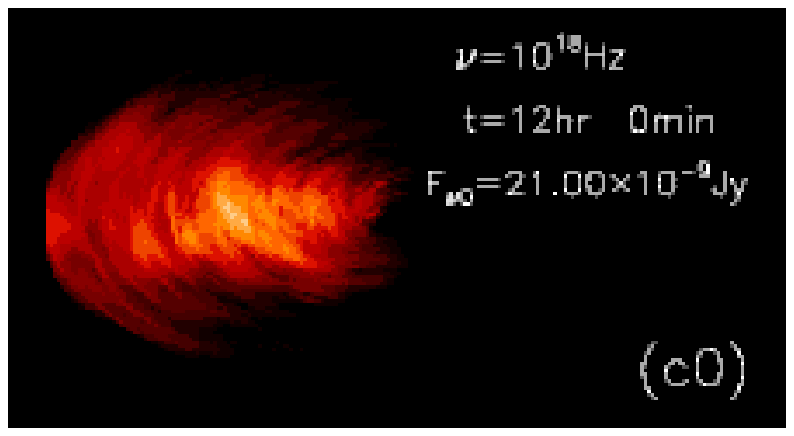} \\
  \includegraphics[scale=0.75]{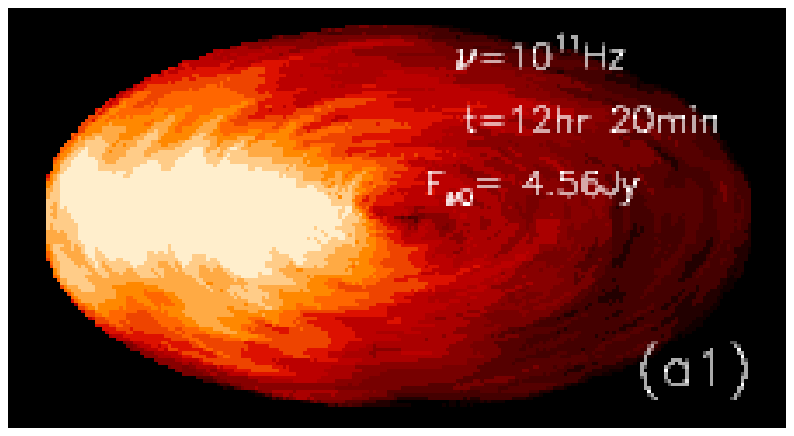}
  \includegraphics[scale=0.75]{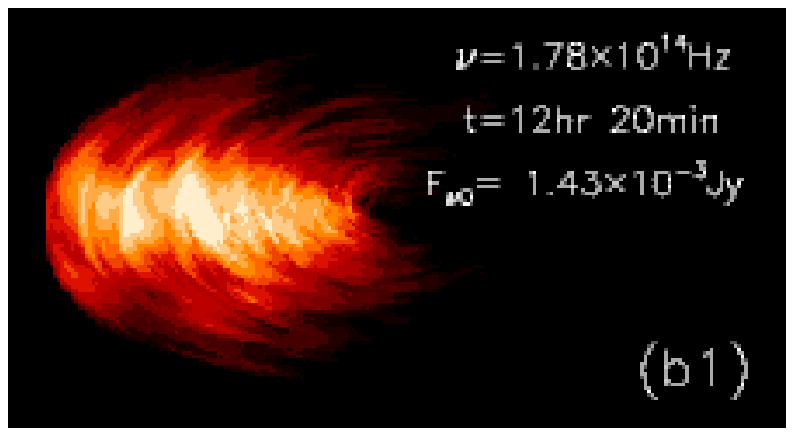}
  \includegraphics[scale=0.75]{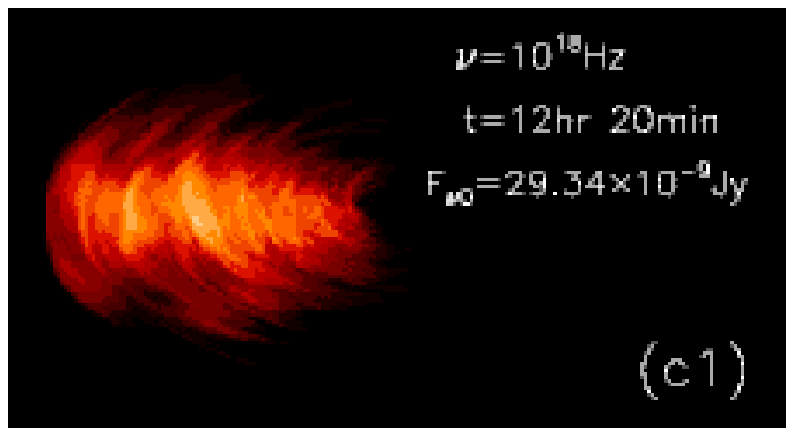} \\
  \includegraphics[scale=0.75]{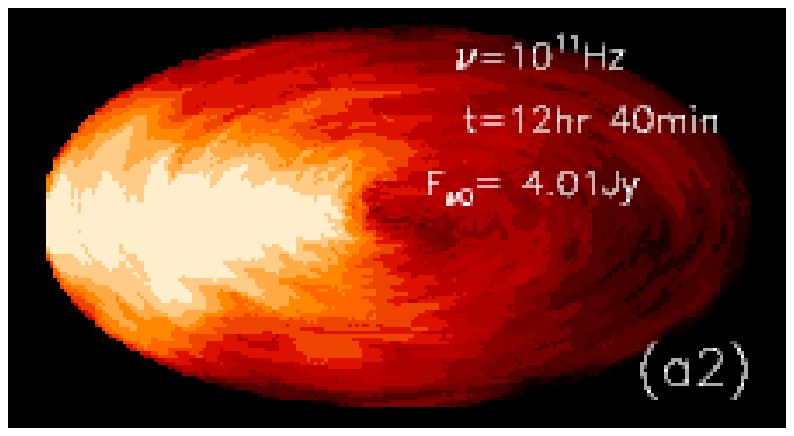}
  \includegraphics[scale=0.75]{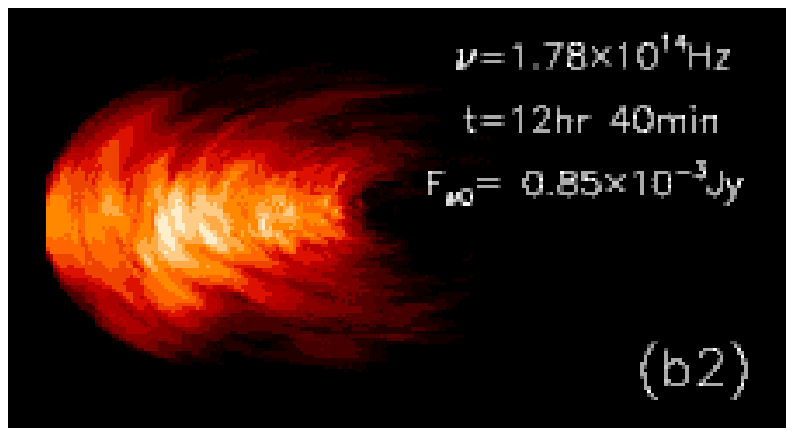}
  \includegraphics[scale=0.75]{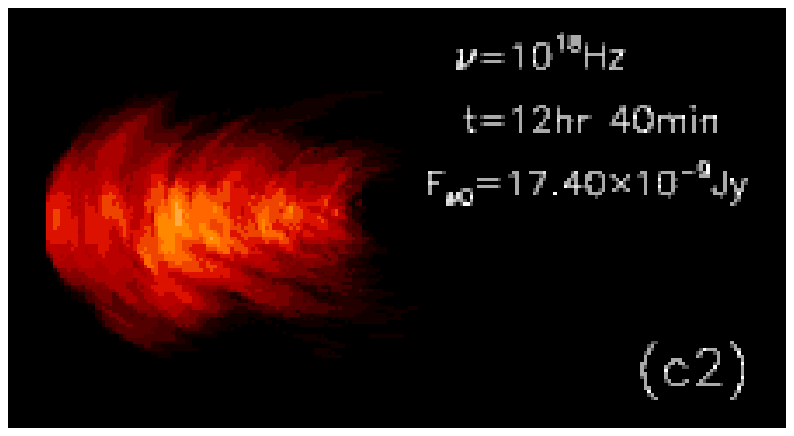} \\
  \includegraphics[scale=0.75]{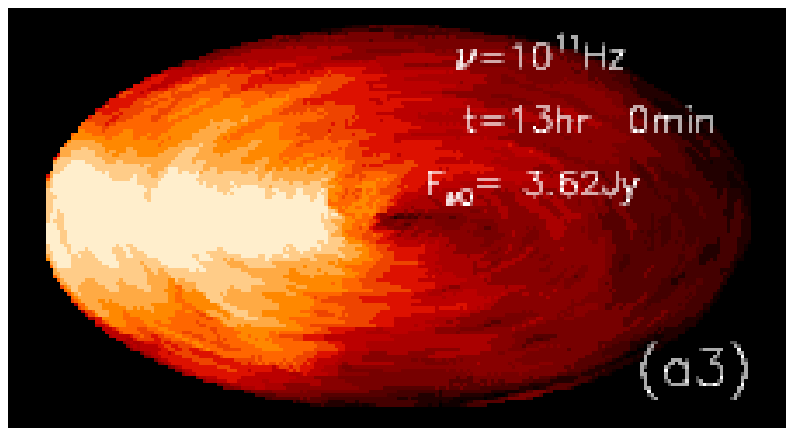}
  \includegraphics[scale=0.75]{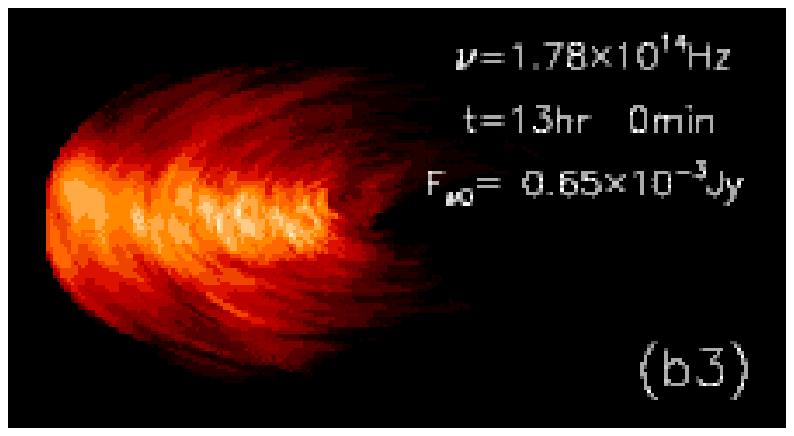}
  \includegraphics[scale=0.75]{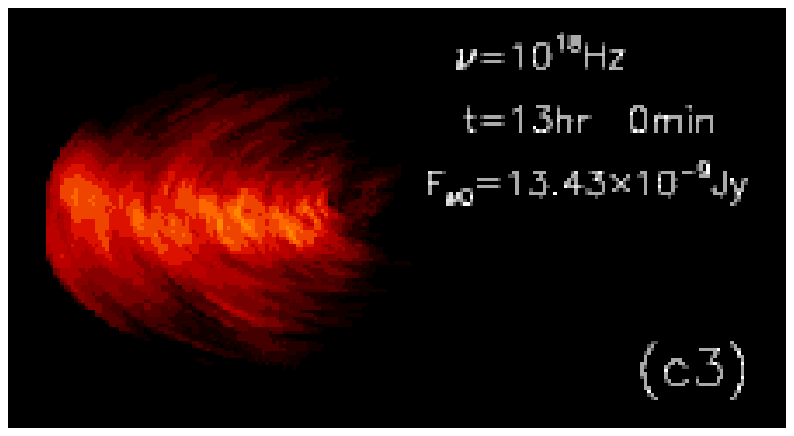} \\
  \includegraphics[scale=0.75]{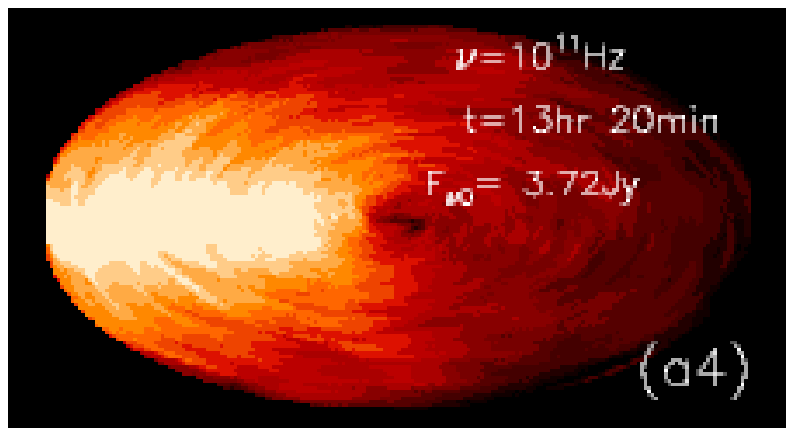}
  \includegraphics[scale=0.75]{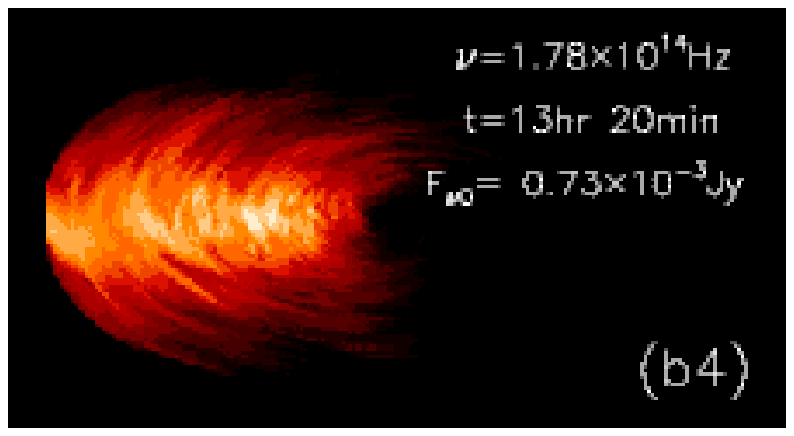}
  \includegraphics[scale=0.75]{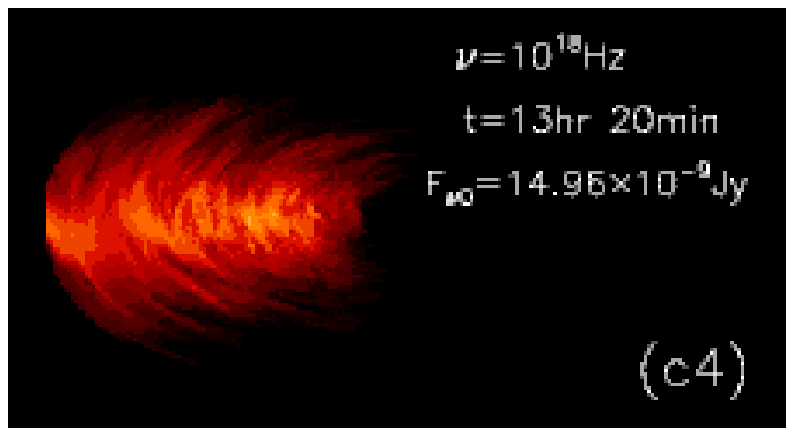} \\
  \includegraphics[scale=0.75]{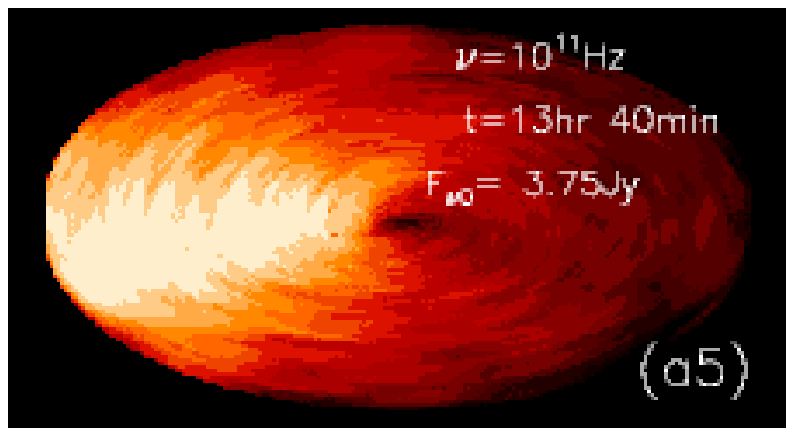}
  \includegraphics[scale=0.75]{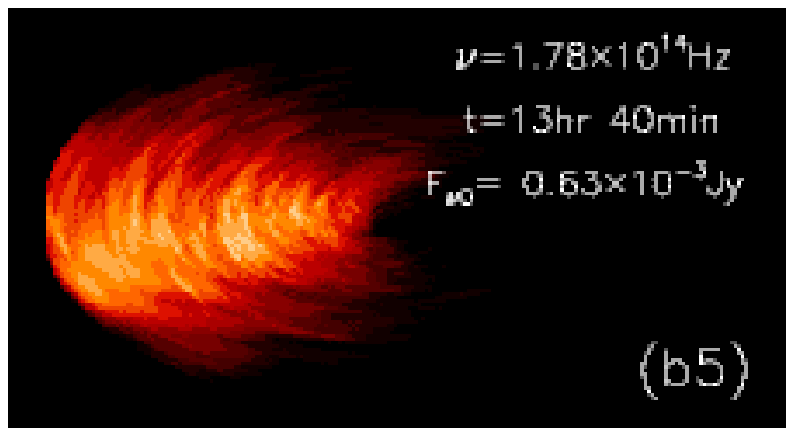}
  \includegraphics[scale=0.75]{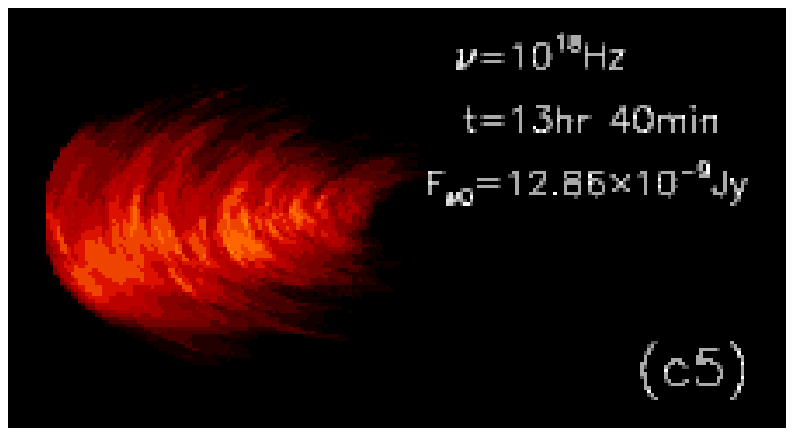} \\
  \includegraphics[scale=0.75]{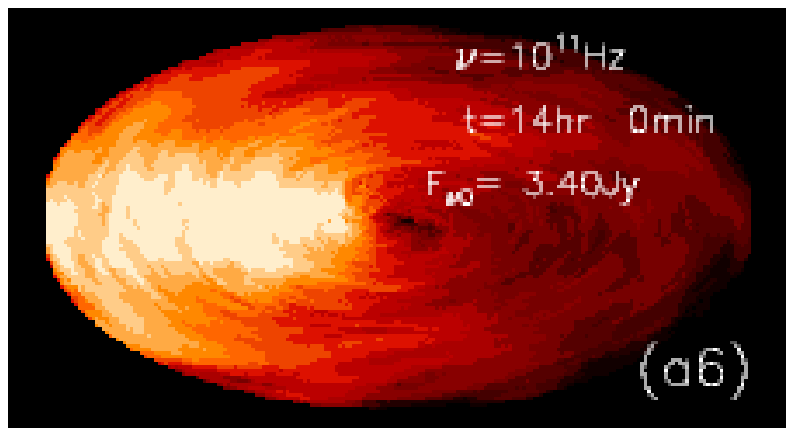}
  \includegraphics[scale=0.75]{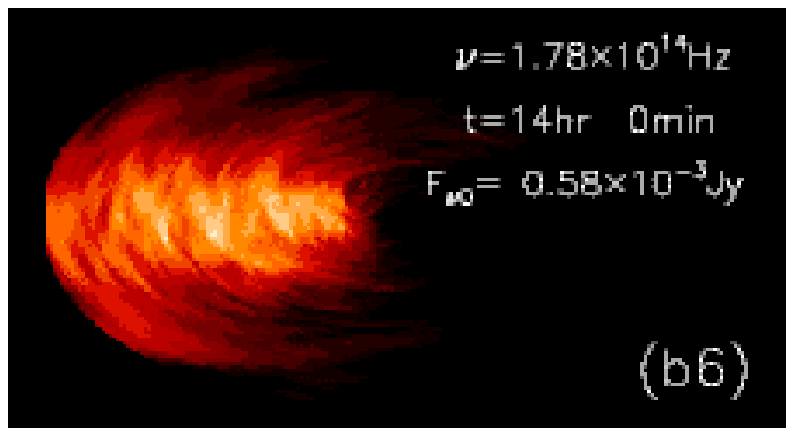}
  \includegraphics[scale=0.75]{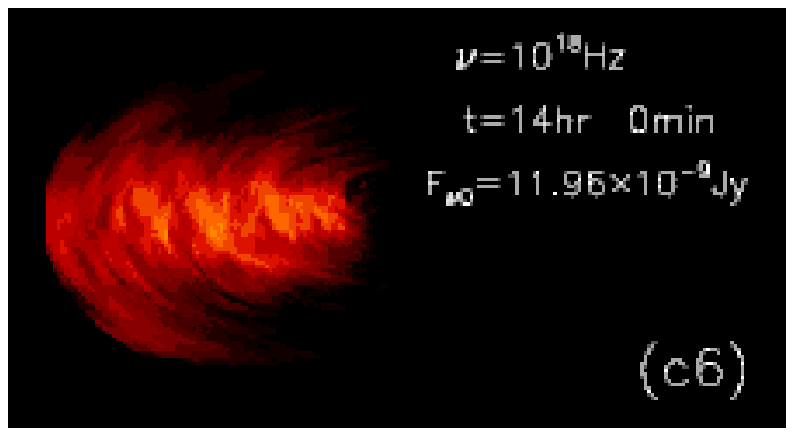}
  \caption{Ray-traced images of the unperturbed simulation of the
    accretion disk in Sgr A*.  The plots show the images at the
    thermal peak (left), infrared (middle), and X-ray (right) bands
    from $t = 12$~hr (top row) to 14~hr (bottom).  The color scale for
    each frequency is fixed so that the reader can see the variation
    of the image with time and compare with Figure~\ref{fig:img_per}
    for the perturbed simulation.  Here, $F_{\nu_0}$ is the flux
    density that would be observed at Earth (from the whole image).
    Note, however, that these images do not include the distortions
    produced by interstellar scattering and the finite telescope
    resolution due to diffraction, and are meant only to convey a
    sense of the emission geometry at the source.}
  \label{fig:img_unp}
\end{figure*}

\begin{figure*}
  \includegraphics[scale=0.75]{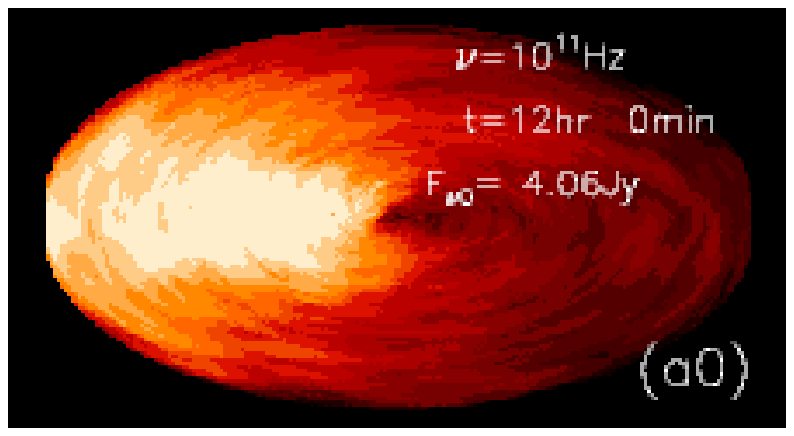}
  \includegraphics[scale=0.75]{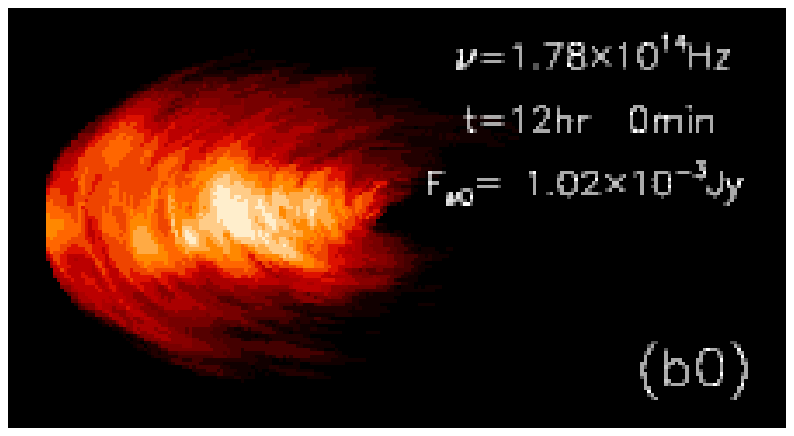}
  \includegraphics[scale=0.75]{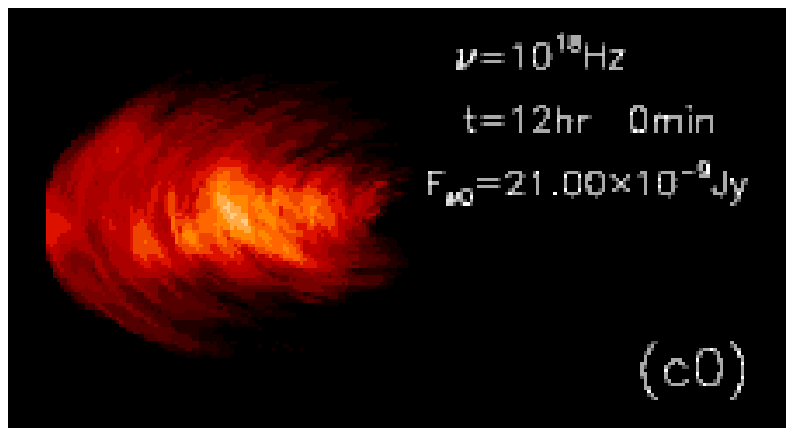} \\
  \includegraphics[scale=0.75]{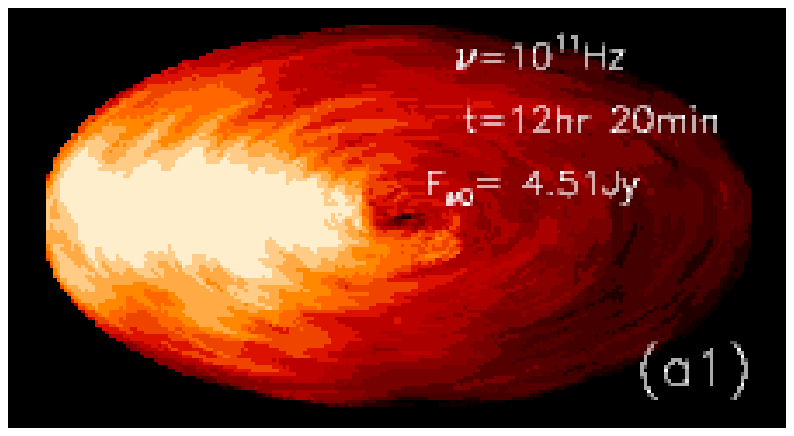}
  \includegraphics[scale=0.75]{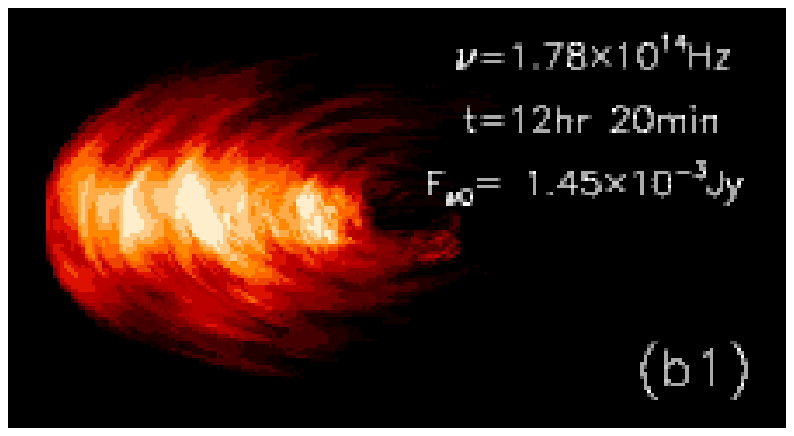}
  \includegraphics[scale=0.75]{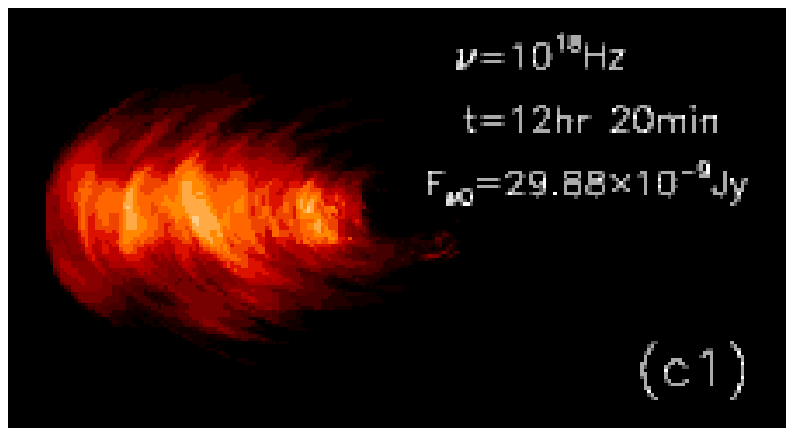} \\
  \includegraphics[scale=0.75]{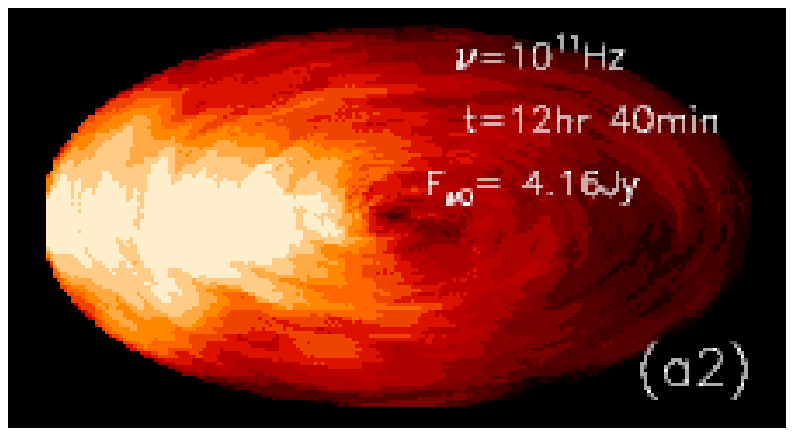}
  \includegraphics[scale=0.75]{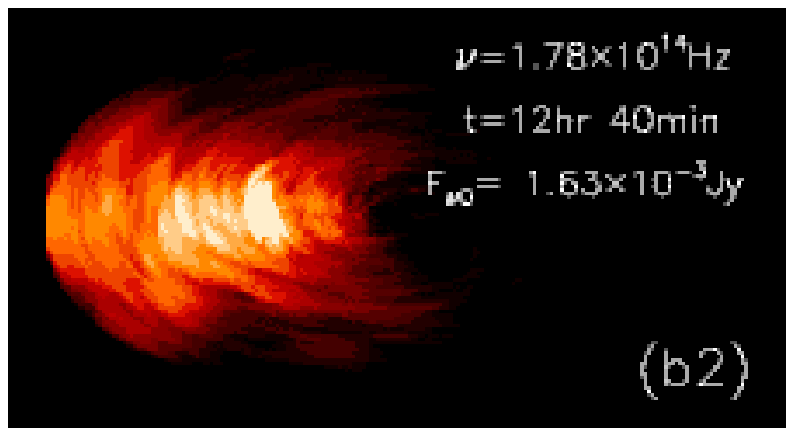}
  \includegraphics[scale=0.75]{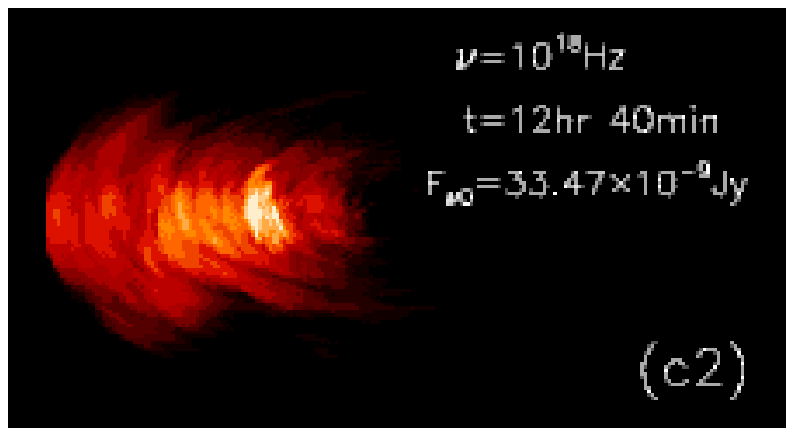} \\
  \includegraphics[scale=0.75]{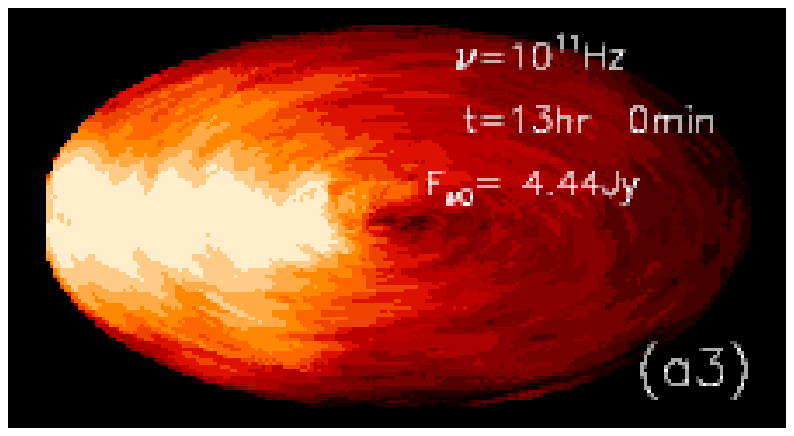}
  \includegraphics[scale=0.75]{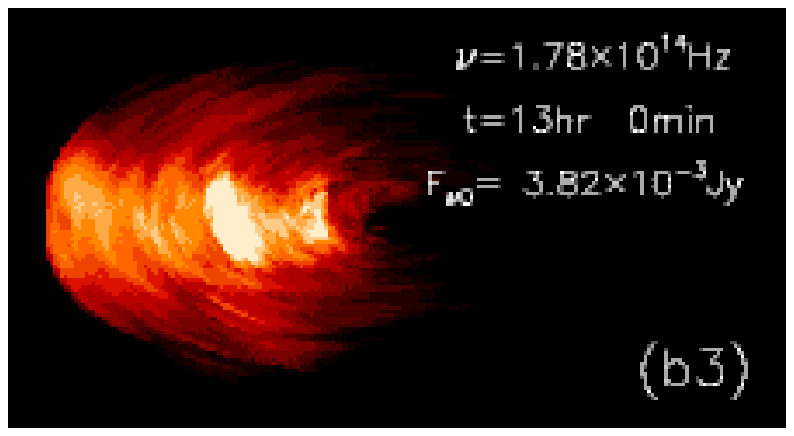}
  \includegraphics[scale=0.75]{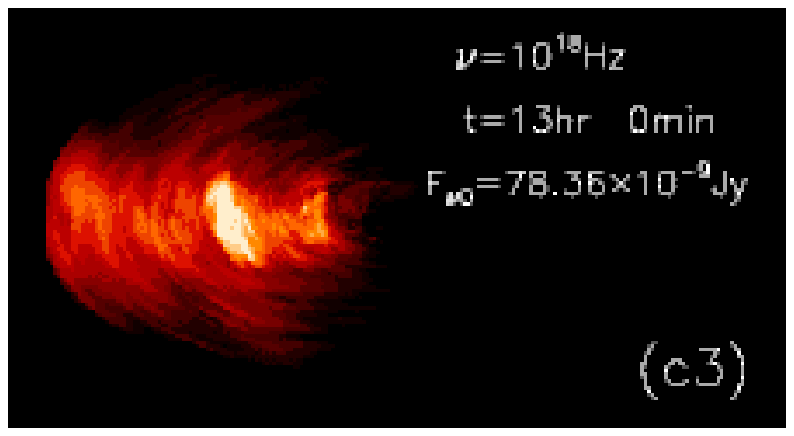} \\
  \includegraphics[scale=0.75]{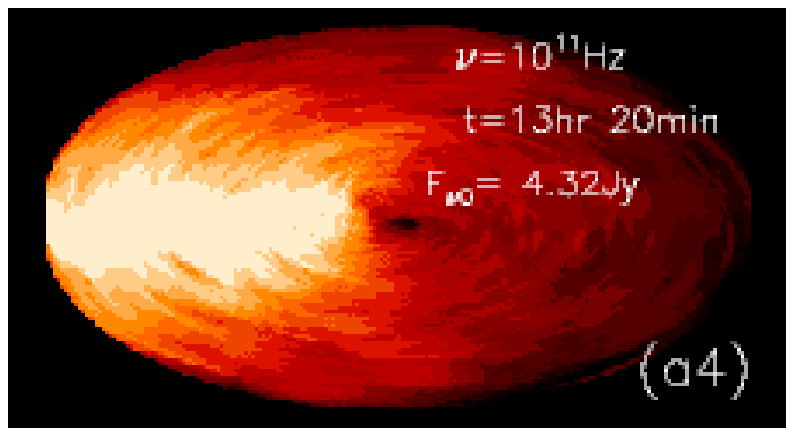}
  \includegraphics[scale=0.75]{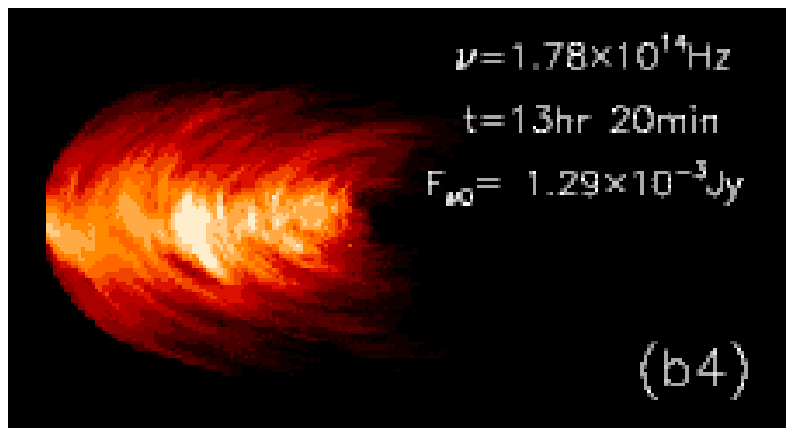}
  \includegraphics[scale=0.75]{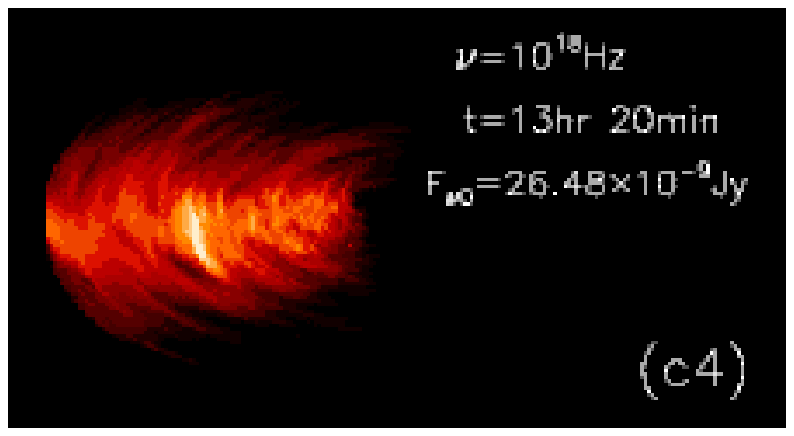} \\
  \includegraphics[scale=0.75]{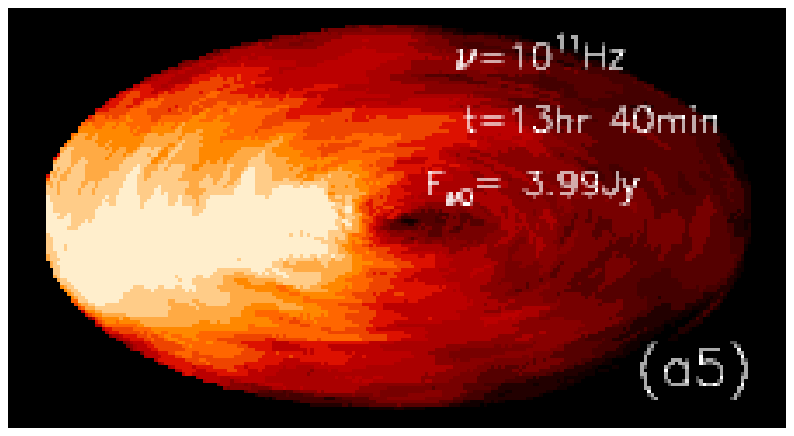}
  \includegraphics[scale=0.75]{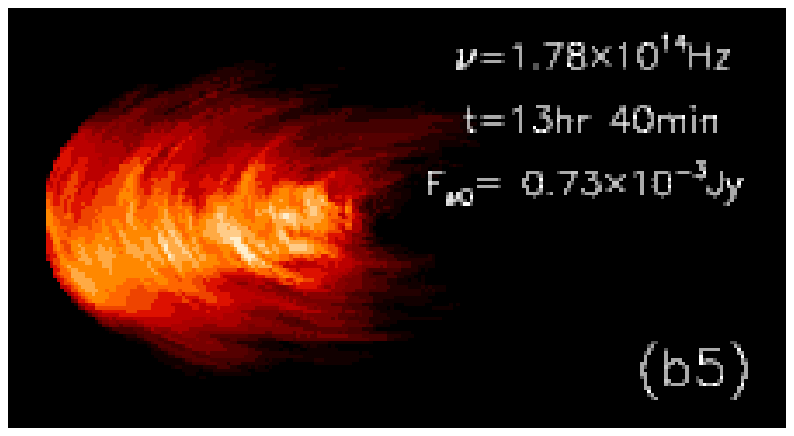}
  \includegraphics[scale=0.75]{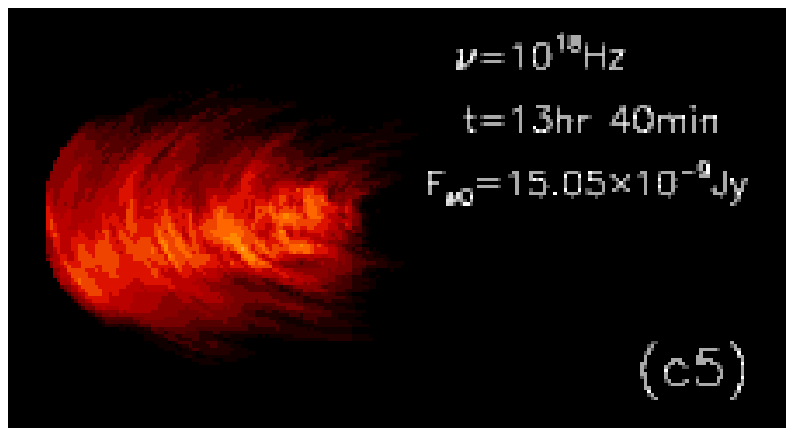} \\
  \includegraphics[scale=0.75]{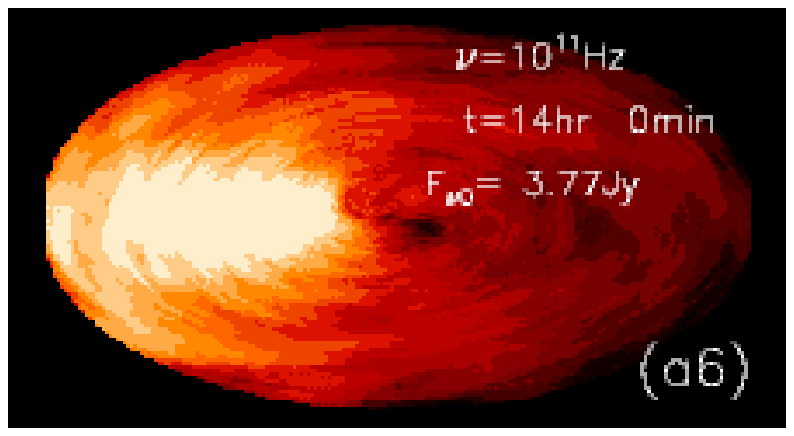}
  \includegraphics[scale=0.75]{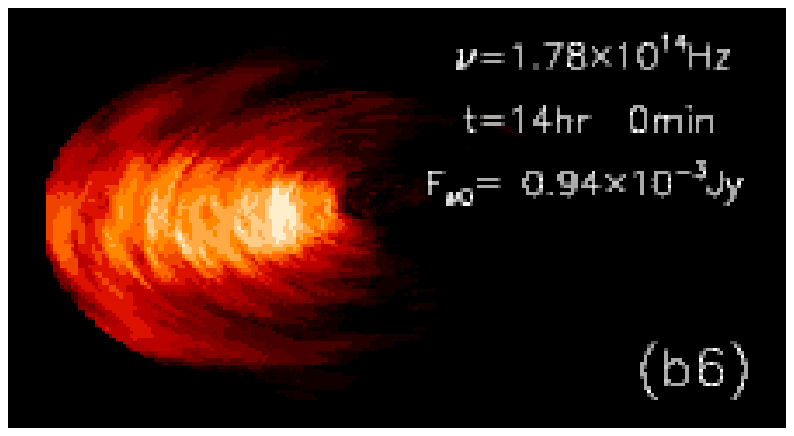}
  \includegraphics[scale=0.75]{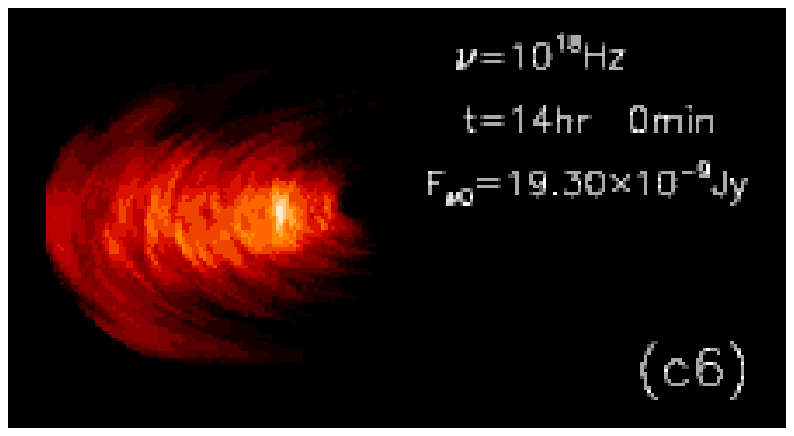}
  \caption{Same as Figure \ref{fig:img_unp} but for the perturbed
    simulation. \\ \\ \\ \\}
  \label{fig:img_per}
\end{figure*}

In Figures~\ref{fig:img_unp} and \ref{fig:img_per} we present the
ray-traced images of our unperturbed and perturbed simulations,
respectively. Note, however, that unlike the images provided in
\citet{Falcke2000} and \citet{Bromley2001}, in which the effects of
interstellar scattering and the diffraction-limited finite resolution
of the telescope array were taken into account to produce realistic
images observed at Earth, these are meant only to show the emission
characteristics at the source.

The images show the disk at mm ($10^{11}$ Hz, left), NIR ($10^{14.25}$
Hz, middle), and X-ray ($10^{18}$ Hz, right) wavelengths, from $t =
12$~hr (top) to 14~hr (bottom). The K-band is centered on 2.2 microns
and has a bandwidth of 0.48 microns. We approximate it by the
frequency $1.78 \times 10^{14} \mathrm{Hz}$, which is equal to
$10^{14.25} \mathrm{Hz}$. The color scales are chosen to enhance the
contrast in the emission and are fixed for each frequency.  The value
of $F_{\nu_0}$ given in the plot is the flux density observed at Earth
from the whole image, according to equation~(\ref{eq:F_nu0}).

From Figure~\ref{fig:img_unp}, we already see that the accretion disk
is highly variable.  The flux densities of the unperturbed disk at
12~hr 20~min and 13~hr 40~min are different by a factor of two in both
the NIR and X-ray bands. They are only different by about 70\% in the
mm/sub-mm bump.  These variations, due to the bright structure at $10
\rS$ [panel (b0)], are consistent with the observed frequency
dependence of the flux variability. This suggests that the turbulent
structure itself can only generate factor-of-two fluctuations.
However, although it is not likely to happen, because we have only run
the simulation for 24~hours, we cannot rule out the possibility that
the turbulence can generate flares that increase the disk brightness
by a larger factor, though much more rarely than these smaller
variations.

The first row of Figure~\ref{fig:img_per} for the perturbed state is
identical to that of Figure~\ref{fig:img_unp} because the perturbation
is added at $\phi = 0$ and $t = 12$~hr.  The flow shears out the
perturbation and forms a one-armed spiral pattern as shown in
Figure~\ref{fig:cont_per}.  Not surprisingly, there is a spiral
pattern in the corresponding images.  At $t = 12\;\mathrm{hr}\
20\mathrm{min}$, we start to see a hot spot appear near the inner edge
of the disk. This hot spot is much brighter than the bright structure
seen in the unperturbed simulation. The hot spot keeps moving outward
due to the \emph{leading} spiral pattern of the perturbation. One hour
after the perturbation is introduced, the disk reaches its brightest
moment when its flux is 20\% higher than its quiescent value in the
radio and 500\% brighter in the NIR and X-ray bands.

In addition to the bright one-armed spiral, there are other bright
features in the inner part of the perturbed disk.  For example, there
are two hot spots in the $t = 13$~hr image.  The inner hot spot moves
much faster than the spiral pattern, and appears in the image about
every half hour after the perturbation is introduced.  These
corotating features suggest that QPOs may be found in these
simulations.

%
%
%

%

\subsection{Light Curves and QPOs}

\begin{figure*}
  \includegraphics[scale=0.75,trim=-9 9 0 18]{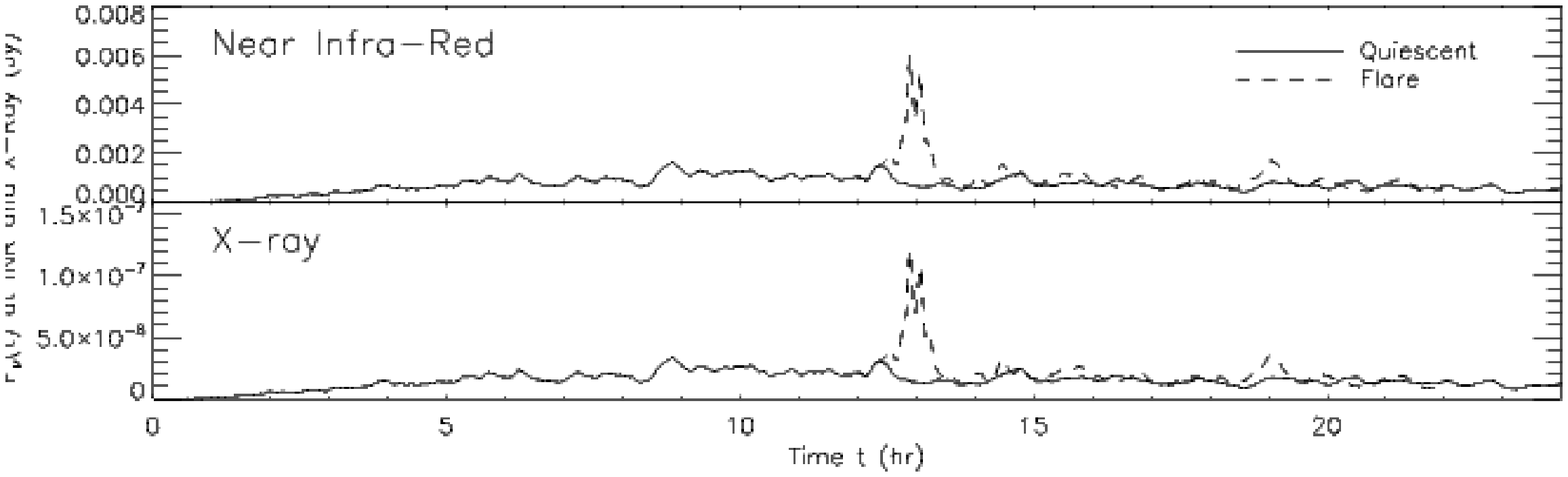}
  \caption{NIR and X-ray lightcurves.  The solid lines are computed
    from the quiescent state, and the dashed lines are from the flare
    simulation. Note that the peaks appearing at the 13th hour are
    much broader than the peak seen in the accretion rate. This
    indicates that the perturbation actually disturbs the
    magnetohydrodynamic disk, instead of trivially raising the
    density. There are some oscillations on top of the peak, which
    persist all the way to the 20th hour. These oscillations
    originating from hot spots in the disk produce a QPO in the
    lightcurve.}
  \label{fig:lc}
  \vspace{6pt}
\end{figure*}

In Figure~\ref{fig:lc}, we show the NIR (upper) and X-ray (lower)
lightcurves obtained by integrating the flux densities over the disk
images.  The solid and dashed lines correspond to the quiescent and
flare state simulations, respectively.

There are several important differences between the lightcurves shown
in Figure~\ref{fig:lc} and the accretion rate plotted in
Figure~\ref{fig:mdot}.  First, note that the initial peak in the
accretion rate in Figure~10 is a transient phenomenon that occurs
because the density relaxes from an initial constant value to a
steady-state solution.  At the same time, the MRI has not yet
amplified the magnetic field to produce significant synchrotron
radiation.  Second, the accretion rate is noisier than the lightcurves
because it is a surface integral at some specific radius (at $3\rS$ in
our case), while the flux density is a volume integral over the whole
computational domain.  Features with a width $\Delta$ in the accretion
disk pass through a specific radius over a time scale $\Delta / v_r$;
however, they can affect the lightcurve for a much longer time
$(r_{\max} - r_{\min})/v_r$.

The flare near $t = 12$~hr in the perturbed lightcurves lasts much
longer than the corresponding peak in the accretion rate.  The NIR
(and X-ray) flux density reaches its peak at $t = 13$~hr, when the
accretion rate at $3.0\rS$ has dropped back to the quiescent level.
These results show that the flux densities are not necessarily
correlated with the accretion rate.  Indeed, our simulations indicate
that the perturbation not only raises the density trivially, but is
also able to change the structure of the turbulence.  This kind of
distortion takes about an hour to develop, and merges with the
original turbulent disk within a second hour.

\begin{figure}[b]
  \includegraphics[scale=0.75,trim=18 9 0 0]{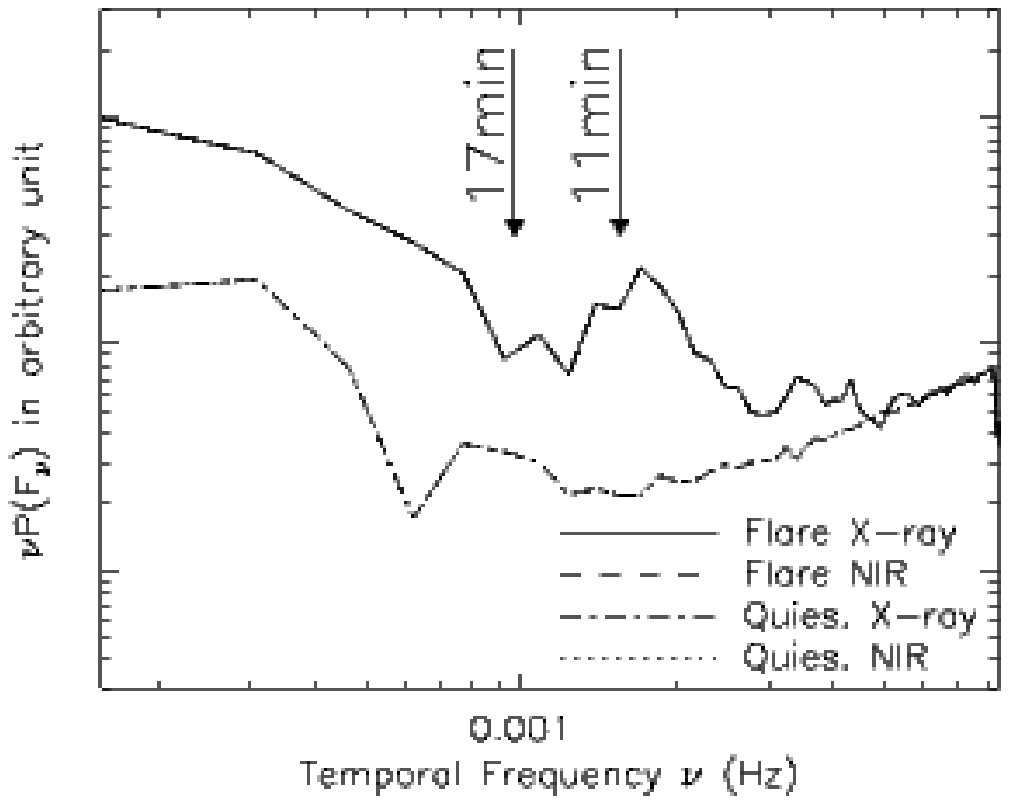}
  \caption{The PSD (multiplied by the frequency) of the lightcurves.
    The curves are normalized so that the flat noise spectra at high
    temporal frequencies match each other.  There is no significant
    period during the quiescent state.  In the perturbed simulation, a
    QPO appears at 11~min, which is the orbital period at $2.4\rS$,
    the magnetosonic point.}
  \label{fig:lc_psd}
\end{figure}

Comparing the lightcurves to the disk images, we can identify the
second peak of the flare with the hot spots shown in panels (b3) and
(c3) in Figure~\ref{fig:img_per}.  It is directly related to the
spiral structure in the disk (see Figure~\ref{fig:cont_per}).
However, the oscillations during the flare, which persist all the way
to about $t = 20$~hr, come from other features developed by the
perturbation---for example, the smaller hot spot located at the inner
boundary in panel (b3) of Figure~\ref{fig:img_per}.

The evolution of these small fluctuations suggests that there should
be a periodic (or quasiperiodic) signal in the lightcurves.  In
Figure~\ref{fig:lc_psd}, we over-plot the power spectral densities
(PSD; multiplied by the frequency) in the NIR band in both the
quiescent and flare state simulations.  The curves are normalized at
higher frequencies for easy comparison of the two. The PSDs of the
X-ray lightcurves are almost identical to (and hence overlapping with)
the NIR PSDs because both of them are produced by the nonthermal
electrons.  The two arrows indicate the frequencies associated with
the orbital period at the ISCO and the magnetosonic point (at $2.4
\rS$).

Although there is no significant period in the quiescent state, a
QPO-like feature is found in the flare simulation.  The quasiperiod is
around 11~min, essentially the Keplerian period at the magnetosonic
point. This is, to our knowledge, the first QPO seen in
magnetohydrodynamic turbulent disks.  If such a density perturbation
(due to ``raining" plasma) is indeed the cause of such flares in
nature, the QPOs observed from Sgr A* may be used to set a \emph{lower
limit} on the orbital period at ISCO, and hence provide an estimate of
the spin of the Galactic center black hole.  This idea can be applied
to other low-luminosity AGN as well.

\section{Limitations}
\label{sec:limitations}

Although for the first time a QPO is seen in magnetohydrodynamic
simulations of accretion disks without invoking special conditions
such as large-scale magnetic fields, the QPO period is shorter than
that observed from Sgr A*, which is usually longer than 17 minutes
\citep{Genzel2003, Belanger2006}.  This may be partially due to the
pseudo-Newtonian potential we have adopted here.  In a Schwarzschild
geometry, the orbital period at the ISCO for a $3.4\times 10^6M_\odot$
black hole is 25.7 minutes.  If the magnetosonic point is still at
$2.4 r_S$ in such a potential, the corresponding period is 18.4
minutes, which is slightly longer than the period observed in the NIR
band but compatible with the X-ray observations \citep{Belanger2006}.
A more complete simulation incorporating general relativistic effects
\citep[e.g.,][]{Hirose2004} and relativistic ray tracing calculations
\citep{Bromley2001} is needed to support more quantitative comparisons
with observations.

We also assume a periodic boundary condition along the vertical
direction and ignore the vertical component of gravity.  Therefore our
simulations are only 2.5-dimensional, in the sense that they are able
to correctly simulate three-dimensional turbulence but cannot capture
many other three-dimensional features.  For example, the geometry does
not allow an outflow along the vertical direction.  Hence, it cannot
be used to determine whether there are jets associated with the hot
accretion disk, which may contribute to nonthermal radio emission from
Sgr A* \citep{Liu2001} and explain the observed correlation between
X-ray flares and radio outbursts \citep{Zhao2004}.  We also cannot see
magnetic-field-dominated funnels as revealed in global relativistic
MHD simulations \citep{Hirose2004}.  These funnels may be responsible
for the acceleration of energetic protons near the black hole
\citep{Liu2006b} and can play an important role in driving
Poynting-flux-dominated outflows, which are proposed to power
large-scale jets in black hole accretion systems.

The saturation level of our magnetic field agrees with the global
simulation of \citet{Hawley2001b, Hawley2002b}.  \citet{Hawley1995}
pointed out that the saturation level of the magnetic field in
shearing box simulations depends on the box size.  The vertical size
of our simulation domain therefore is a free parameter.
\citet{Stone1996} also showed that there are more complicated
structures in the vertical direction when the effects of gravity are
included.  A fully three-dimensional simulation is needed to resolve
these issues.

To compare observations with MHD simulations, one also has to convert
the simulated disk characteristics into a radiation flux spectrum.
Observations have already provided good constraints on the emission
processes.  However, it is not clear how electrons are energized in
the turbulent plasma, which is related to the fundamental problem of
non-linear turbulence dissipation and electron-ion coupling.  In the
case of relativistic collisionless plasmas considered here, we lack
even the appropriate tools to handle this subject self-consistently.
There have been only some phenomenological models developed to address
this problem quantitatively \citep{Liu2004, Liu2006a, Bittner2006}.
In this paper, we have adopted one of the simplest emission models
appropriate for the accretion flow in Sgr A* and have used
observations to constrain the model parameters.  Although the QPO
signal is not very sensitive to the model parameters as long as there
is a higher energy electron population above the thermal background,
one should exercise caution with more quantitative comparisons.

Besides the points mentioned above, there are additional physical
processes we can include to improve our simulations and quantitative
results.  For example, thermal bremsstrahlung, and inverse Compton
scattering of the synchrotron photons by the hot electrons may provide
a significant contribution to the X-ray emission \citep{Narayan1998}
especially during big flares \citep{Liu2004}.

\section{Conclusions}
\label{sec:conclusions}

In this paper we have simulated a nonradiative MHD accretion disk
around Sgr A* using a pseudo-spectral algorithm.  The results are
consistent with earlier three-dimensional MHD simulations of the MRI
\citep{Hawley2002a}, even though we have assumed a different geometry.
Our simulations reached a quasi-steady state after the MRI-driven
turbulence has fully developed.

We computed the long-wavelength spectrum from our simulation by using
a hybrid emission scheme with a thermal background of electrons plus a
high-energy nonthermal broken power-law tail.  We then used the
broadband quiescent-state spectrum to calibrate the model parameters.
In the quasi-steady state we found that the source is variable with an
amplitude in flux density that increases with photon frequency, in
agreement with observations. However, the flux density varied by only
a factor two, failing to account for the observed NIR and X-ray flares
from Sgr A*.

Motivated by earlier studies of the large-scale accretion processes in
Sgr A* \citep[e.g.,][]{Coker1997, Cuadra2005}, we introduced a density
perturbation to the quasi-steady state disk to see whether they could
produce even bigger flux density variations.  We found that bigger
flares were produced in both the NIR and X-ray bands, as well as a QPO
in the lightcurves in association with the magnetosonic point below
the ISCO.  The fact that the simulated QPO is associated with the
magnetosonic point below the ISCO also requires a reexamination of
previous estimates of the black hole spin based on flare observations
\citep{Genzel2003, Aschenbach2004}.

\acknowledgments

This work was carried out under the auspices of the National Nuclear
Security Administration of the U.S. Department of Energy at Los Alamos
National Laboratory under Contract No. DE-AC52-06NA25396.  At the
University of Arizona, this research was supported by NSF grant
AST-0402502 and NASA ATP grant NAG5-13374, and has made use of NASA's
Astrophysics Data System Abstract Service.  F.M. is grateful to the
University of Melbourne for its support (through a Sir Thomas Lyle
Fellowship and a Miegunyah Fellowship). The simulations were carried
out on the Space Simulator at Los Alamos National Laboratory and on a
Beowulf cluster in the Physics Department at the University of
Arizona.


\end{document}